\documentclass[submission, Phys]{SciPost}
\usepackage{amssymb}
\usepackage{amsmath}
\usepackage{mathrsfs}  
\usepackage{subcaption} % in preamble
\usepackage{placeins}
\usepackage{booktabs}

\hypersetup{
    colorlinks,
    linkcolor={red!50!black},
    citecolor={blue!50!black},
    urlcolor={blue!80!black}
}

\usepackage{tikz}
\usetikzlibrary{arrows.meta, positioning, shadows.blur, backgrounds, fit}

\usepackage[bitstream-charter]{mathdesign}
\DeclareSymbolFont{usualmathcal}{OMS}{cmsy}{m}{n}
\DeclareSymbolFontAlphabet{\mathcal}{usualmathcal}

\begin{document}

\begin{center}{\Large \textbf{
    A ``Neural'' Riemann solver for Relativistic Hydrodynamics
}}\end{center}

% TODO: write the author list here. Use first name (+ other initials) + surname format.
% Separate subsequent authors by a comma, omit comma and use "and" for the last author.
% Mark the corresponding author with a superscript star.
\begin{center}
Carlo Musolino\textsuperscript{1$\star$}
\end{center}

\begin{center}
{\bf 1} Institute for Theoretical Physics,\\ Max-von-Laue Str. 1, 60438, Frankfurt am Main, Germany
\\
${}^\star$ {\small \sf musolino@itp.uni-frankfurt.de}
\end{center}

\begin{center}
\today
\end{center}

% For convenience during refereeing (optional),
% you can turn on line numbers by uncommenting the next line:
%\linenumbers
% You should run LaTeX twice in order for the line numbers to appear. 8 lines

\section*{Abstract}
{\bf
In this paper, we present an approach to solving the Riemann problem in one-dimensional relativistic hydrodynamics, where the most computationally expensive steps of the exact solver are replaced by compact, highly specialized neural networks. The resulting ``neural'' Riemann solver is integrated into a high-resolution shock-capturing scheme and tested on a range of canonical problems, demonstrating both robustness and efficiency. By constraining the learned components to the root-finding of single-valued functions, the method retains physical interpretability while significantly accelerating the computation. The solver is shown to achieve accuracies comparable to the exact algorithm at a fraction of the cost, suggesting that this approach may offer a viable path toward more efficient Riemann solvers for use in large-scale numerical relativity simulations of astrophysical systems.
}

\vspace{10pt}
\noindent\rule{\textwidth}{1pt}
\tableofcontents\thispagestyle{fancy}
\noindent\rule{\textwidth}{1pt}
\vspace{10pt}

\section{Introduction}
\label{sec:intro}
At the heart of modern finite-volume methods for solving hyperbolic conservation laws lies the Riemann problem — the initial value problem with piecewise constant data separated by a single discontinuity. This problem needs to be solved---either exactly or via an approximation---at every cell interface to reconcile the fluxes coming from data at either side into a single numerical flux~\cite{Godunov59,Marti96,colella_1984_ppm}. 

In the context of relativistic hydrodynamics, exact solutions to the Riemann problem are known and can be computed to arbitrary precision~\cite{Marti94,Pons00,Rezzolla01}. However, their computational cost renders them impractical for use in high-resolution shock-capturing (HRSC) codes. As a result, most production codes rely on approximate Riemann solvers. Among these, one of the most widely adopted is the HLLE solver — originally proposed by Harten, Lax, and van Leer, and later extended by Einfeldt in the context of classical gas dynamics~\cite{Harten1976,Einfeldt88}.

In the HLLE method, the complex wave structure resulting from the decay of the initial discontinuity is replaced by a single intermediate state, which approximates an integral average of the exact Riemann fan. Numerical fluxes are then computed using the Rankine-Hugoniot jump conditions, ensuring that the scheme remains consistent with the conservation laws. First introduced to relativistic hydrodynamics in~\cite{Schneider93,Duncan93a}, this approach is widely appreciated for being computationally efficient, simple to implement, and consistent by construction with physical constraints such as positivity and entropy conditions.

Despite its popularity, the HLLE Riemann solver is known to suffer from certain limitations. Its simplicity comes at the cost of neglecting important features of the Riemann solution, leading to excessive diffusion of sharp discontinuities such as contact surfaces and shear waves. A common remedy to this issue is the HLLC Riemann solver~\cite{Toro1994,Mignone2005,Kiuchi2022a}, which introduces an additional intermediate state separated by an approximate contact discontinuity. The propagation speed of this discontinuity is typically estimated based on the HLLE solution. While this refinement reduces some of the artificial diffusion associated with HLLE, it still fails to capture the full richness of wave interactions inherent in the relativistic Riemann problem.

This naturally raises the question: \textit{can further improvements in Riemann solver accuracy lead to better-resolved features in HRSC codes, without compromising computational efficiency?}

Within the context of these premises, Machine Learning methods constitute a natural avenue to explore. The use of neural networks in scientific research has gained significant traction in recent years. Variants of deep neural networks have been successfully applied to the reconstruction of unknown differential operators from noisy data~\cite{Rackauckas2011,Lu2021}, to the solution of ordinary and partial differential equations~\cite{Raissi2019}, as well as to the discovery of physical laws from empirical observations~\cite{Udrescu2020,Udrescu2020b}. Similar methods have also recently been employed in the context of relativistic hydrodynamics to accelerate the solution of the conservative to primitive inversion~\cite{Kacmaz2025,Mudimadugula:2025piz}, a well known bottleneck in simulation codes~\cite{Galeazzi2013,Kastaun2021}. Neural networks have been applied in various contexts to solve the Riemann problem either in isolation or embedded within Godunov-like schemes~\cite{Wang2023,Patsatzis2024,Ruggeri2022,Magiera2020,Gyrya2024}. Related efforts also include predicting the full solution of hyperbolic PDEs from discontinuous initial data~\cite{Peyvan2024}, following an operator-learning approach similar to DeepONets~\cite{Lu2021}. Most of these methods rely on monolithic network architectures that either predict fluxes or interfacial states directly, or attempt to mimic classical approximate solvers by enforcing the Rankine-Hugoniot conditions. To the best of our knowledge, such strategies have not yet been extended to the equations of relativistic hydrodynamics.

With this in mind, we explore a different direction where instead of replacing the entire solver with learned components, we leverage well-scoped, compact neural networks trained on exact solutions to the Riemann problem, aimed at circumventing the main computational bottlenecks that make traditional exact solvers impractical for large-scale simulations. More concretely, the method presented in this paper employs a collection of simple feed-forward neural networks, each conditioned on a single wave pattern, to replace the computationally expensive iterative steps in the traditional exact solver. This structure can be viewed as a physics-informed Mixture of Experts (MoE)~\cite{Shazeer2017}, where analytical criteria are used to route each Riemann problem to a specialized sub-network based on the underlying wave pattern. In doing so, we demonstrate that data-driven Riemann solvers can combine high physical fidelity with competitive runtime performance, offering a promising path toward next-generation approximate solvers.

This work is organised as follows: Sec.~\ref{sec:hydro} presents the formalism and the basic equations of relativistic hydrodynamics, Sec.~\ref{sec:riemann} presents the Riemann problem and its solution, Sec.~\ref{sec:neural_solver} presents the method, Sec.~\ref{sec:training} describes the architecture and training strategies of the networks employed in this work, Sec.~\ref{sec:results} showcases the solver's accuracy and performance in a series of real-world simulations one-dimensional testcases and Sec.~\ref{sec:future} highligths some potential avenues for future developments of this method. Finally, we draw our conclusions in Sec.~\ref{sec:conclusions}.
\section{Relativistic hydrodynamics} \label{sec:hydro}
The motion of a relativistic compressible fluid is governed by the conservation of mass, energy and momentum. The corresponding equations can be cast in flux-conservative form, which in $3+1$ dimensions read:
\begin{equation}\label{eq:rhd_cons}
    \partial_t U_{a} + \partial_i F_{a}^i\left( U_{b} \right) = 0\, ,
\end{equation}
where the index $i$ runs over spatial coordinates from $1$ to $3$, and the indices $a,b$ distinguish between different state variables. In Eq.~\eqref{eq:rhd_cons}, the vector of conserved variables $U_a$ is defined as~\cite{Marti99,Banyuls97}
\begin{equation} \label{eq:cons}
    U_{a} := 
    \begin{pmatrix} 
        D \\ 
        S^i \\
        \tau
    \end{pmatrix}
    =
    \begin{pmatrix} 
        W \rho \\ 
        W^2 \rho h v^i \\ 
        W \, \rho ( W h - 1 ) - p  
    \end{pmatrix}\, ,
\end{equation}
where $v^i$ is the fluid's three-velocity, $W=(1-v^i\,v_i)^{-1/2}$ is the Lorentz factor, $h$ is the specific enthalpy, $\rho$ the rest-mass density and $p$ the pressure. Note that we follow the convention in which the evolved variable \(\tau := E - D\) represents the total energy density (excluding rest mass) which is widely adopted in general relativistic hydrodynamics codes used in astrophysical contexts~\cite{Moesta2013,Radice2012a,Radice2015,Most2019b}. In turn, the flux vector $F^i_a$ is given by 
\begin{equation} \label{eq:fluxes}
    F^i_a = 
    \begin{pmatrix} 
        D \, v^i \\
        S_j v^i + p \, \delta^i_j \\ 
        S^i - D\, v^i
    \end{pmatrix}\, ,
\end{equation}
where $\delta^i_j$ is the Kronecker delta. For convenience, let us introduce the vector of primitive variables which is defined to be 
\begin{equation} \label{eq:prims}
    P_a = 
    \begin{pmatrix} 
        \rho \\ 
        p \\
        v^i
    \end{pmatrix} \, . 
\end{equation}
 The set of Eqs.~\eqref{eq:cons}---relating primitive variables to conservatives ones---is closed by an equation of state (EOS), i.e.~a relation giving the enthalpy in terms of other thermodynamic quantities $h=h(\rho, p)$ (or equivalently, the pressure in terms of rest-mass density and specific internal energy, $\epsilon$). To simplify the presentation, the analysis in this work is restricted to the ideal gas equation of state---often referred to as $\gamma$ law---whereby~\cite{Rezzolla_book:2013}
\begin{equation}\label{eq:eos}
    h(\rho,p) = 1 + \frac{\gamma \, p }{(\gamma-1)\, \rho}\, .
\end{equation}
For future convenience, we report here the expression of the sound speed $c_s$ for a gamma-law EOS
\begin{equation} \label{eq:cs}
    c_s^2 = \gamma\,\frac{p}{\rho\, h}\, .
\end{equation}
\section{The Riemann problem}\label{sec:riemann}
The Riemann problem for relativistic hydrodynamics has been studied in detail by several authors, and there exist algorithms to determine its solution---both in the one as well as in multi-dimensional cases~\cite{Marti94,Pons00,Rezzolla01}---to arbitrary accuracy. This solution will consist of three self-similar wave patterns arising from the decay of the initial discontinuity. The resulting Riemann fan will be formed by two nonlinear waves propagating from the initial jump at speeds $\lambda_{\rm R/L}$, separated by a contact discontinuity moving at a speed $\lambda_{\rm C}$. The main complication in solving the Riemann problem boils down to determining the states of the fluid on either side of the contact discontinuity. This can be achieved by making use of the fact that, across said discontinuity, the normal component of the velocity and the pressure are continuous, leading to the following nonlinear equation
\begin{equation} \label{eq:riemann_rootfind}
    v^*_{\rm R}(p^*) = v^*_{\rm L}(p^*)\, , 
\end{equation}
where the analytic form of the functions $v^*_{\rm L/R}(p^*)$ depends on the character of the left/right going nonlinear wave. We will now give a working summary of the main results and expressions relating to the wave patterns that can arise from a Riemann problem in relativistic hydrodynamics in $1+1$ dimensions, and refer the interested reader to reviews on the topic for more detailed accounts~\cite{Rezzolla_book:2013,Marti94,Pons00}. 
\subsection{Rarefaction waves}
A rarefaction wave is a nonlinear wave characterized by the fact that both the density and the pressure decrease in the region behind the wave's head. They are isentropic solutions of the flow equations, and they can be written in self-similar form with respect to the variable $\xi=x/t$. Due to this fact, it can be shown that the sound speed within the rarefaction wave obeys 
\begin{equation} \label{eq:cs_xi}
    c_s = \pm \frac{v - \xi}{1 - v \xi}\, ,
\end{equation}
where the $+$ (resp. $-$) sign refers to a left (resp right) going rarefaction wave. Because of the isentropic, self-similar nature of the flow through a rarefaction wave, the equations of fluid motion can be expressed as an exact differential 
\begin{equation}
    W^2 dv \pm \frac{c_s}{\rho} d\rho = 0 \, , 
\end{equation} 
which implies that the Riemann invariant 
\begin{equation} \label{eq:jpm}
    \mathcal{J}_\pm = \frac{1}{2} \log\left( \frac{1+v}{1-v} \right) \pm \int \frac{c_s}{\rho} d \rho \, ,
\end{equation}
is constant.
In the case of an ideal gas EOS Eq.~\eqref{eq:jpm} can be integrated analytically, allowing us to find the required expression relating the velocity and the pressure behind a rarefaction wave 
\begin{equation} \label{eq:raref}
    v^*_{\rm L/R} = \frac{(1+v_{\rm L/R})\, A_\pm (p^*) - (1-v_{\rm L/R})}{(1+v_{\rm L/R})\, A_\pm (p^*) + (1-v_{\rm L/R})} \, ,
\end{equation}
where $A_+$ and $A_-$ correspond to the left and right going rarefaction waves and are given by 
\begin{equation}
    A_\pm (p) = \left\{ \left[ \frac{(\gamma - 1 )^{\frac{1}{2}} - c_s(p) }{(\gamma - 1 )^{\frac{1}{2}} + c_s(p) } \right]  \left[ \frac{(\gamma - 1 )^{\frac{1}{2}} + c_s(p_{\rm L/R}) }{(\gamma - 1 )^{\frac{1}{2}} - c_s(p_{\rm L/R}) } \right] \right\}^{\pm 2/(\gamma-1)^{\frac{1}{2}}} \, .
\end{equation}
\subsection{Shockwaves}
Shockwaves are discontinuous solutions of the flow equations, characterized by an increase in entropy of the fluid going through the wave's surface. The integral form of the equations across the discontinuity leads to the so-called Rankine-Hugoniot jump conditions, which can be used to relate the fluid variables ahead and behind the shock. In particular, the integral form of the rest-mass conservation equation across a hypersurface where the fluid variables are discontinuous can be stated as 
\begin{equation}\label{eq:rankine_one}
    [\![j]\!] = 0 \, ,
\end{equation}
where, as is customary, $[\![X]\!]$ stands for the difference of a quantity $X$ across the discontinuity, and 
\begin{equation}\label{eq:j_shock}
    j = W_s \, D \,( V_s - v ) \, ,
\end{equation}
where $V_s$ is the propagation speed of the shock and $W_s$ the corresponding Lorentz factor. It can be shown that the rest of the equations of motion lead to 
\begin{equation} \label{eq:rankine_two}
    j^2 = -\frac{[\![p]\!]}{[\![h/\rho]\!]} \, ,
\end{equation}
and 
\begin{equation} \label{eq:taub}
    [\![h^2]\!] = \left( \frac{h_b}{\rho_b} + \frac{h_a}{\rho_a} \right)\;[\![p]\!]  \, ,
\end{equation}
the latter being commonly referred to as the Taub adiabat, where subscripts $a$ and $b$ indicate quantities ahead of and behind the shock, respectively. 
Eqs.~\eqref{eq:rankine_two} ,~\eqref{eq:taub}, and~\eqref{eq:rankine_one}, together with the EOS, form a system of four equations for the five unknowns $(V_s, \rho_b, p_b, h_b, v_b)$. This is already sufficient to obtain the required function relating the post-shock velocity $v_b$ to the corresponding pressure $p_b$. Despite this fact, it is useful to note that---in the case of an ideal gas EOS---the Taub adiabat can be rewritten explicitly as a quadratic polynomial in $h_b$ that does not depend on $\rho_b$
\begin{equation}
 \left( 1 + \frac{(\gamma-1)(p_a-p_b)}{\gamma\,p_b}\right)\,h_b^2 - \frac{(\gamma-1)(p_a-p_b)}{\gamma\,p_b} \, h_b + \frac{h_a\,(p_a-p_b)}{\rho_a} - h_a^2 = 0 \, ,
\end{equation}
which makes the solution of the system considerably easier.
\subsection{Exact solution of the Riemann problem}
Equipped with Eq.~\eqref{eq:raref} and the correspoding equation for shockwaves, all that is needed is a prescription to determine, given the initial data, the wave pattern that will arise in the Riemann fan. 
In this regard,~\cite{Rezzolla01} found that---analogously to the non-relativistic case---the wave pattern corresponding to a given set of discontinuous initial data can be determined by the invariant relative velocity between the two states alone. In particular, since the expressions for post-wave velocities are monotonic, limiting values that distinguish among the three possible wave patterns: \textit{(i)} a double shock ($2{\rm S}$), \textit{(ii)} a double rarefaction ($2{\rm R}$), or \textit{(iii)} a mixed shock-rarefaction (${\rm S}{\rm R}$) can be obtained analytically. 

In summary, the solution of the Riemann problem can be broken down in the following steps: 
\begin{enumerate}
    \item Determine the wave pattern by comparing the relative velocity $\Delta v_{\rm LR} = (v_{\rm L} - v_{\rm R})/(1 - v_{\rm L} v_{\rm R})$ to the limiting values of~\cite{Rezzolla01}
    \item Find the states on either side of the contact discontinuity by solving Eq.~\eqref{eq:riemann_rootfind}
    \item Solve for the rest of the fluid variables by using the EOS and wave-specific expressions.
\end{enumerate}

Once the primitive variables on either side of the contact discontinuity ($P^*_{\rm L}$ and $P^*_{\rm R}$) have been determined, the numerical flux at the interface can be computed by evaluating the solution at $\xi = 0$. The value of this flux depends on which region of the self-similar Riemann fan contains the origin. Specifically:

\begin{equation}
    F^*_{\rm exact} = 
    \begin{cases}
    F_{\rm L}      & \text{if}~ \lambda_{\rm L} \geq 0 \, ,\\
    F_{\rm RL}     & \text{if}~ \lambda_{\rm L} < 0 ~ \text{and}~ \xi_{\rm L} \geq 0 \, ,\\
    F_{\rm CL}     & \text{if}~ \xi_{\rm L} < 0 ~ \text{and}~ \lambda_{\rm C} \geq 0 \, ,\\
    F_{\rm CR}     & \text{if}~ \lambda_{\rm C} < 0 ~ \text{and}~ \xi_{\rm R} \geq 0 \, ,\\
    F_{\rm RR}     & \text{if}~ \xi_{\rm R} < 0 ~ \text{and}~ \lambda_{\rm R} \geq 0 \, ,\\
    F_{\rm R}      & \text{if}~ \lambda_{\rm R} \leq 0 \, .
    \end{cases}
\end{equation}

Here, $F_{\rm L}$ and $F_{\rm R}$ are the fluxes computed from the initial states on the left and right of the interface, respectively. The fluxes $F_{\rm CL}$ and $F_{\rm CR}$ are computed from the primitive variables just to the left and right of the contact discontinuity. The fluxes $F_{\rm RL}$ and $F_{\rm RR}$, on the other hand, correspond to the rarefaction fan regions when the wave is expanding away from the interface. In such cases, the state at $\xi = 0$ lies inside the rarefaction wave, and its primitive variables cannot be obtained directly from the post-wave state. Instead, they must be computed by integrating the rarefaction wave structure using the Riemann invariant defined in Eq.~\eqref{eq:jpm}, which ultimately boils down to the solution of a scalar nonlinear equation\footnote{In the case of higher-dimensional flows this is an ODE, due to the coupling of velocity components via the Lorentz factor.}. These intermediate states play a crucial role in constructing the full solution and are also predicted by neural networks in the data-driven solver presented here.

In practice, this procedure forms the foundation of exact Riemann solvers for relativistic hydrodynamics, which are widely used for code validation and benchmarking. However---as mentioned in the Introduction---the nonlinear root-finding step in Eq.~\eqref{eq:riemann_rootfind} (and optionally in Eq.~\eqref{eq:jpm}, if the interface lies within a rarefaction wave) makes exact solvers prohibitively expensive for large-scale simulations. Instead, modern numerical codes typically rely on approximate Riemann solvers, such as HLLE or HLLC~\cite{Mignone2005}, which bypass the full solution in favor of reduced models.

While a Godunov-type scheme employing any Riemann solver that satisfies the Rankine-Hugoniot jump conditions and basic physical consistency will converge (in the weak sense) to the correct solution of the PDE, in practice, the choice of solver can significantly impact the resolution of physical features at finite grid spacing. Improved Riemann solvers can capture contact discontinuities and shock fronts more sharply, reducing numerical diffusion and enhancing the accuracy of the simulation~\cite{Mignone2005,Kiuchi2022a}.

With this in mind, we will now describe our strategy towards the development of a data-driven Riemann solver that mirrors the classical solution strategy---determining the exact states within the Riemann fan---while circumventing the need for expensive root-finding. This is achieved by replacing the nonlinear inversion with appropriately trained neural networks that directly predict the fluid properties at cell interfaces.
\section{``Neural'' Riemann solver}\label{sec:neural_solver}

As stated in the previous section, the goal is to replace the computational bottlenecks in the exact Riemann solver with a data-driven alternative that reproduces the solution up to high accuracy, but at a fraction of the cost. To achieve this, this work adopts a hybrid strategy: we retain the analytic structure of the solution, including wave pattern classification and flux selection logic, but replace the most expensive parts with neural networks trained to approximate the corresponding map. This hybrid structure allows us to retain the physical interpretability and consistency of the solver while accelerating its performance.

More concretely, the neural solver involves the following learned components:
\begin{itemize}
\item Three pattern-specific neural networks $g_{\text{p}}$ --- where $p\in \{ {\rm 2S}, {\rm 2R}, {\rm SR} \}$ stands for double shock, double rarefaction, and shock rarefaction --- trained to predict the pressure $p^*$ across the contact discontinuity from the left and right primitive states. These networks replace the root-finding step in Eq.~\eqref{eq:riemann_rootfind}.

\item Two direction-specific rarefaction state networks $r_{\leftarrow}$ and $r_\rightarrow$,  used to reconstruct the primitive variables at the interface ($\xi = 0$) when it lies inside the rarefaction fan. This step is required to compute the fluxes $F_{\rm RL}$ or $F_{\rm RR}$ depending on the direction of the rarefaction.
\end{itemize}
\noindent After the pressure---or, in the case of rarefactions, the sound speed---at the interface has been predicted by the appropriate network, the remaining variables are computed analytically using wave-specific relations. This confines the uncertainty inherent to neural inference and ensures that the underlying analytic structure of the Riemann problem is utilized to the fullest. This hybrid strategy has also proven effective in previous works involving other PDE systems~\cite{Magiera2020,Ruggeri2022}.

Each of these networks can be viewed as learning an implicitly defined map 
\begin{equation} \label{eq:solution_map}
g\colon\; \theta \mapsto x^*\, ,
\end{equation}
where $x^*$ is the solution to a nonlinear equation of the form 
\begin{equation} \label{eq:rootfinding_general}
    f(x^*, \theta) = 0\, .
\end{equation}
For instance, in the case of the contact pressure solver, $\theta$ consists of the left and right primitive variables and $x^*=p^*$ is the pressure that equalizes the normal velocities across the contact discontinuity. Similarly, for rarefaction reconstruction, the network learns the mapping from left/right region variables to the sound speed at $\xi=0$ within the rarefaction fan.

Under mild regularity conditions---which are satisfied throughout all physically admissible regimes---the Implicit Function Theorem guarantees the local existence and differentiability of these solution maps. Moreover, since the relevant input domains are well-behaved for each wave pattern, the Universal Approximation Theorem\cite{Hornik1991} ensures that feed-forward neural networks can approximate these maps to arbitrary precision.\footnote{Note that these differentiability conditions depend on the functional properties of the equation of state, which enters the solution of both shock and rarefaction branches.}

Importantly, we emphasize that each wave pattern is handled by a separate neural network. This design choice not only aligns with the natural decomposition of the Riemann problem but also ensures that each network approximates a smooth, single-valued function---avoiding the complexity of representing discontinuous or multi-branched behaviors within a single architecture. This strategy differs from most prior attempts to learn the full solution manifold directly and is central to the effectiveness and stability of the solver.

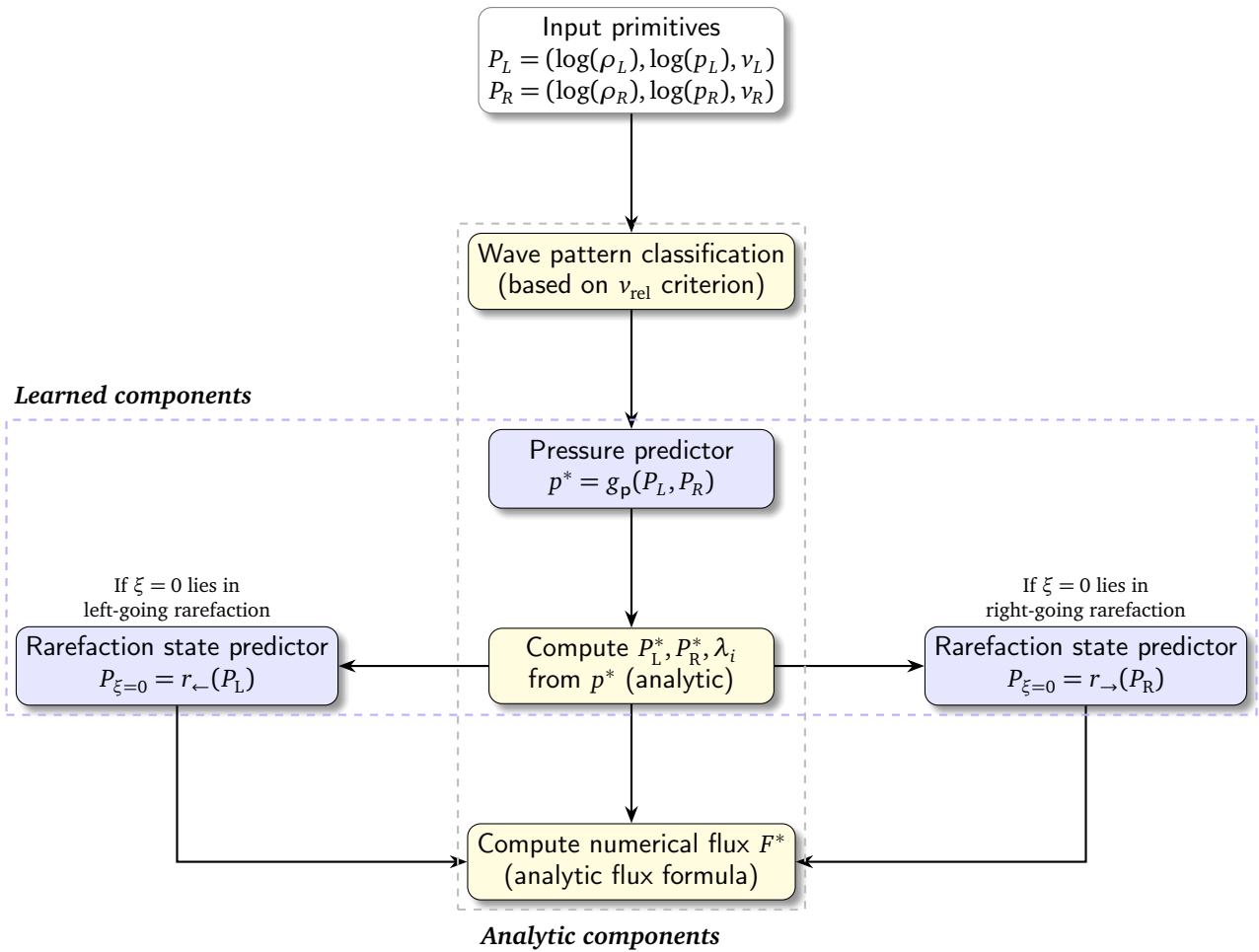
\begin{figure}[htbp]
    \centering
    \makebox[\textwidth][c]{
        \begin{tikzpicture}[
            node distance=1.6cm and 2cm,
            every node/.style={font=\sffamily\small},
            process/.style={rectangle, draw, rounded corners=5pt, align=center, minimum width=3.8cm, minimum height=1cm, blur shadow},
            learned/.style={process, fill=blue!10},
            logic/.style={process, fill=yellow!15},
            io/.style={process, fill=white, draw=gray},
            arrow/.style={-{Stealth}, thick}
          ]
          
          % Input
          \node[io] (input) {Input primitives\\$P_L = (\log(\rho_L),\log(p_L), v_L)$\\$P_R = (\log(\rho_R),\log(p_R), v_R)$};
          
          % Pattern classifier
          \node[logic, below=of input] (pattern) {Wave pattern classification\\(based on $v_{\rm rel}$ criterion)};
          
          % Pressure net
          \node[learned, below=of pattern] (pnet) {Pressure predictor\\$p^* = g_{\text{p}}(P_L, P_R)$};
          
          % Recover P*
          \node[logic, below=of pnet] (recoverP) {Compute $P^*_{\rm L}, P^*_{\rm R}, \lambda_i$\\from $p^*$ (analytic)};
          
          % Rarefaction nets
          \node[learned, right=of recoverP] (rnetR) {Rarefaction state predictor\\$P_{\xi=0} = r_{\rightarrow}(P_{\rm R})$};
          \node[learned, left=of recoverP] (rnetL) {Rarefaction state predictor\\$P_{\xi=0} = r_{\leftarrow}(P_{\rm L})$};
          
          % Compute flux
          \node[logic, below=of recoverP] (flux) {Compute numerical flux $F^*$\\(analytic flux formula)};
        
          % Arrows
          \draw[arrow] (input) -- (pattern);
          \draw[arrow] (pattern) -- (pnet);
          \draw[arrow] (pnet) -- (recoverP);
          \draw[arrow] (recoverP) -- (flux);
          \draw[arrow] (recoverP.east) -- ++(0.8,0) -- (rnetR.west);
          \draw[arrow] (rnetR.south) |- (flux.east);
          \draw[arrow] (recoverP.west) -- ++(-0.8,0) -- (rnetL.east);
          \draw[arrow] (rnetL.south) |- (flux.west);

          % Conditions (moved below rarefaction nets for clarity)
            \node[draw=none, align=center, font=\scriptsize, anchor=north] 
            at ($(rnetR.north) + (0,+0.8)$) {If $\xi = 0$ lies in \\ right-going rarefaction};

            \node[draw=none, align=center, font=\scriptsize, anchor=north] 
            at ($(rnetL.north) + (0,+0.8)$) {If $\xi = 0$ lies in \\ left-going rarefaction};
        
          % Group boxes
          \node[draw=blue!30, dashed, thick, fit=(pnet)(rnetL)(rnetR)] (learnedbox) {};
          \node[draw=gray!50, dashed, thick, fit=(pattern)(recoverP)(flux)] (analyticbox) {};

          \node[draw=none, align=center, font=\small\itshape, anchor=south west] at ($(learnedbox.north west) + (0.,0.)$) {\textbf{Learned components}};

          \node[draw=none, align=center, font=\small\itshape, anchor=south west] at ($(analyticbox.south west) + (0.2,-0.65)$) {\textbf{Analytic components}};
        \end{tikzpicture}
    }
\caption{Schematic architecture of the neural Riemann solver. Neural networks are used to predict the contact pressure and, when necessary, the state within rarefaction fans. Remaining steps follow analytic expressions.}\label{fig:neural_solver_diagram}
\end{figure}

A schematic overview of the learned solver architecture is shown in Fig.~\ref{fig:neural_solver_diagram}.

\section{Network architecture and training}\label{sec:training}
As detailed in the previous section, the approach of this paper builds on the traditional analytic solution to the Riemann problem, replacing only those steps which are too computationally expensive to perform \textit{online} with neural networks. This hybrid strategy retains the interpretability and structure of the exact solver, while allowing for significant acceleration. Moreover, since the networks are small and well-scoped, physical insights from the underlying equations can be used to constrain their output, improving both robustness and generalizability.

All networks used in this work follow a shared architecture: a fully connected feed-forward multilayer perceptron (MLP) with a few (typically two) hidden layers of fixed width. Two core physical requirements must be respected to ensure the overall scheme remains physically viable:
\begin{enumerate}
    \item \textbf{Positivity}: The reconstructed states within the Riemann fan must satisfy basic physical constraints — namely, positive pressure and rest-mass density, and subluminal velocities. In conservative variables, this translates to the requirement that $D > 0$ and $E > \sqrt{m^2 + D^2}$.
    \item \textbf{Physical consistency}: Godunov-type schemes converge (in the weak sense) to the correct solution of hyperbolic conservation laws only if the numerical fluxes obey the Rankine-Hugoniot jump conditions. These conditions ensure consistency with shock propagation and entropy production.
\end{enumerate}
It is instructive to recall how these conditions are enforced in a classical approximate solver such as HLLE.\@ There, the intermediate state is computed as an integral average over a spacetime control volume that spans the Riemann fan, ensuring positivity by construction. The flux is then computed directly from a discretized form of the Rankine-Hugoniot relations --- rather than by evaluating $F(U^*)$ --- which ensures that condition (2) is satisfied.

In the case of the neural Riemann solver, condition (1) is also enforced by construction: the network output is scaled and passed through a sigmoid function to guarantee it remains within a positive, bounded interval. Since the other primitive variables are recovered analytically from the predicted quantity ($p^*$ in the case of the $g_{\rm p}$ nets, $c_s$ in the case of $r_\leftarrow$ and $r_\rightarrow$), enforcing positivity on this value ensures that the resulting state is physical. 

Condition (2), on the other hand, is not satisfied exactly---the predicted root will not exactly solve the target equation, but will approximate it to within the network's error. This deviation from exact Rankine-Hugoniot satisfaction is the key tradeoff introduced by the present method, and one that is explicitly controlled during training.

To this end, let us define the network output as:
\begin{equation}\label{eq:net_out}
    \chi = \phi(\beta\,\mathcal{N}(\theta))\, ,
\end{equation}
where $\theta$ denotes the input vector of fluid variables, which is rescaled to the range $[-1,1]$ using fixed minimum and maximum values. These bounds are the same as those used during synthetic data generation (see Appendix~\ref{appendix:data_generation} for details). The function $\mathcal{N}(\theta)$ represents the raw output of the feed-forward network, $\phi$ is a sigmoid activation mapping to the interval $(0,1)$, and $\beta$ is a learnable scaling parameter. The scalar $\chi$ is then mapped to the corresponding physical output range via a network-specific transformation.

Each network is trained by minimizing a loss function that blends two objectives:
\begin{equation} \label{eq:loss}
\mathcal{L}_\theta(x) = \mathcal{H}(f(x,\theta), 0) + \mathcal{H}(\log (x),\log (x^*))\, ,
\end{equation}
where $f(x,\theta)$ is just the constituent relation of~\eqref{eq:rootfinding_general}---i.e.~the residual of the nonlinear constraint being solved---$x^*$ is the exact root obtained offline, and $\mathcal{H}$ is the Huber loss~\cite{Huber64}. The first term enforces the physics-based constraint, while the second encourages accurate recovery of the correct root. The use of the Huber loss helps mitigate the influence of outliers and prevents instability from large gradient spikes during training~\cite{Girshick2015}.

We will now briefly describe the architecture and training of each of the networks used in this work. The detailed description of the algorithms used to generate training data can be found in Appendix~\ref{appendix:data_generation}.
\subsection*{Contact pressure predictors}
These networks predict the pressure $p^*$ across the contact discontinuity for each of the three possible wave patterns. The input consists of the six-dimensional state $P_L, P_R$, encoded as $\log(\rho), \log(p), v$ for each side. The output $\chi$ is mapped to the physical range of $p^*$ using:
\begin{itemize}
    \item An affine mapping between $\text{min}(p_L,p_R)$ and $\text{max}(p_L,p_R)$ for $g_{\rm SR}$.
    \item An affine mapping between a small number $\epsilon$\footnote{In the tests $\epsilon$ is always set to $10^{-45}$. This is done because---while all expressions have well behaved limits for $p=0$ analytically---numerically they can lead to round-off cancellations and instabilities.} and $\text{min}(p_L,p_R)$ for $g_{\rm 2R}$.
    \item A compactification between $\text{max}(p_L,p_R)$ and $+\infty$ for $g_{\rm 2S}$.
\end{itemize}
The training data is generated using a quasi-random Sobol sequence to uniformly sample the high-dimensional input space. Specifically, the relative velocity between initial states is restricted to lie within the range corresponding to the target wave pattern---computed from the analytic limiting velocities described in Sec.~\ref{sec:riemann}. This ensures that the training data for each network is physically consistent with its intended wave configuration.
\subsection*{Rarefaction fan predictors}
These networks predict the primitive state at $\xi = 0$ when the interface lies within a rarefaction fan. The input is a three-dimensional primitive state (from the side the rarefaction originates). The output is mapped to the sound speed $c_s(\xi=0)$ using a sigmoid scaled to the physical range $(0,c_s(p_{\rm R/L}))$, while the other quantities are reconstructed analytically using the isentropic relation.

In this case the training data is also generated using a quasi-random Sobol sequence in the three-dimensional input space, and further filtered to avoid unphysical configurations. The full logic and implementation are described in Appendix~\ref{appendix:data_generation}.

All models are trained for $100$ epochs using the Adam optimizer~\cite{Kingma2014} with an initial learning rate of $10^{-2}$. The dataset---consisting of $2^{17}$ samples for each network---is randomly split into $80\%$ training and $20\%$ validation sets. Batches of $128$ entries are loaded with random shuffling. The learning rate is halved whenever the validation loss plateaus for more than 5 epochs~\cite{Bengio2012}. To mitigate instabilities, we apply gradient clipping with a maximum norm of $1.0$~\cite{Pascanu2013}, and mix the root and residual losses using the PCGrad strategy to avoid conflicting gradients~\cite{Yu20}. The training behaviour of the neural model is showcased in Appendix~\ref{appendix:training}.
\section{Results}\label{sec:results} 
We now present the results of a series of one-dimensional relativistic hydrodynamics simulations designed to assess the performance of the neural Riemann solver. We will compare both its accuracy and computational efficiency against three standard approaches: the exact Riemann solver and the approximate HLLE and HLLC solvers.

The numerical algorithm used to evolve the system is based on the semidiscrete form of the conservation equations~\eqref{eq:rhd_cons}, specialized to one spatial dimension: 
\begin{equation} 
    \frac{\partial U_a}{\partial t} = \frac{1}{\Delta x} \, \left( F^a_{i-1/2} - F^a_{i+1/2}\right) \, , 
\end{equation} 
where $i$ labels the spatial cell index, and the interfacial fluxes $F_{i \pm 1/2}$ are computed using a Riemann solver applied to left and right states reconstructed from cell averages.

Two reconstruction strategies are considered: 
\begin{enumerate} 
    \item \textit{Flat reconstruction}: Following the original Godunov scheme~\cite{Godunov59}, the interfacial states are defined as $U_{i + 1/2, {\rm L}} = U_i$ and $U_{i + 1/2, {\rm R}} = U_{i+1}$, resulting in a linear scheme that is limited to first-order accuracy by Godunov's Theorem. 
    \item \textit{WENO reconstruction}: Third order weighted essentially non-oscillatory polynomial reconstruction~\cite{Liu1994,Jiang1996,Shu99} is used to compute interfacial states. This reconstruction allows the overall scheme to retain second-order convergence in smooth regions of the flow (note that the flux differences are computed with a central stencil of that order).
\end{enumerate}

Time integration is performed using the Method of Lines~\cite{Hyman-1976-Courant-MOL-report} (MoL) with either the first-order forward Euler method or the second-order midpoint rule (a variant of Runge-Kutta) and the conservative-to-primitive inversion is performed using the method of~\cite{Galeazzi2013}. Our implementation is built using the \texttt{PyTorch} library~\cite{PyTorch2}. Notably, the use of the \texttt{torchdiffeq} package~\cite{torchdiffeq} makes the entire hydrodynamics solver fully differentiable. While this feature is not exploited in this work, it may enable future extensions involving differentiable programming or hybrid training strategies.

When using the exact and neural Riemann solvers, it is necessary to prescribe a treatment for the case in which the left and right primitive states at a cell interface are nearly identical. Strictly speaking, continuous data does not constitute a genuine Riemann problem, and in particular, when $\Delta v_{\rm LR} = 0$, the resulting wave pattern would be identified as a shock-rarefaction. However, explicitly evaluating the post-shock expressions in this regime can lead to numerical issues due to round-off error when the pre- and post-shock pressures are nearly equal. To ensure robustness, both the exact and neural solvers fall back to the HLLE flux when the absolute difference between all reconstructed primitive variables on either side of the interface is less than a predefined threshold $\epsilon_c$, which we fix to $10^{-10}$ throughout this work.

Moreover, to reduce the number of possible wave patterns and simplify the logic of both the exact and neural solvers, we exploit the symmetry of the Riemann problem by reflecting states whenever $p_{\rm R} > p_{\rm L}$, so that all inputs satisfy $p_{\rm L} \geq p_{\rm R}$. In the neural solver, this reduces the degeneracy of the target function learned by each network by explicitly enforcing symmetry under reflection. The corresponding fluxes are then reflected back to restore the correct orientation before the conservative update is applied.

All tests described in this Section are conducted using the same neural solver---i.e., a single collection of networks trained without any problem-specific fine-tuning of the training data. The neural architecture consists of two hidden layers with $64$ neurons for the pressure predictors and $32$ neurons for the rarefaction solvers. 

All sources of randomness involved in training---including weight initialization, dataset shuffling and batching, and the random engine used for dataset generation---are controlled using the same fixed seed ($42$). The only exception is Fig.\ref{fig:l1_fo}, where results from $11$ different seeds are shown to illustrate variability. A full analysis of the solver's sensitivity to the random seed and other hyperparameters is provided in Appendix~\ref{appendix:ablation}.

All simulations are performed using an ideal gas equation of state with adiabatic index $\gamma = 5/3$, and a Courant-Friedrichs-Lewy (CFL) factor of $0.8$.
\subsection{Problem 1}
\begin{figure}[htbp]
    \centering
    \includegraphics[width=0.8\textwidth]{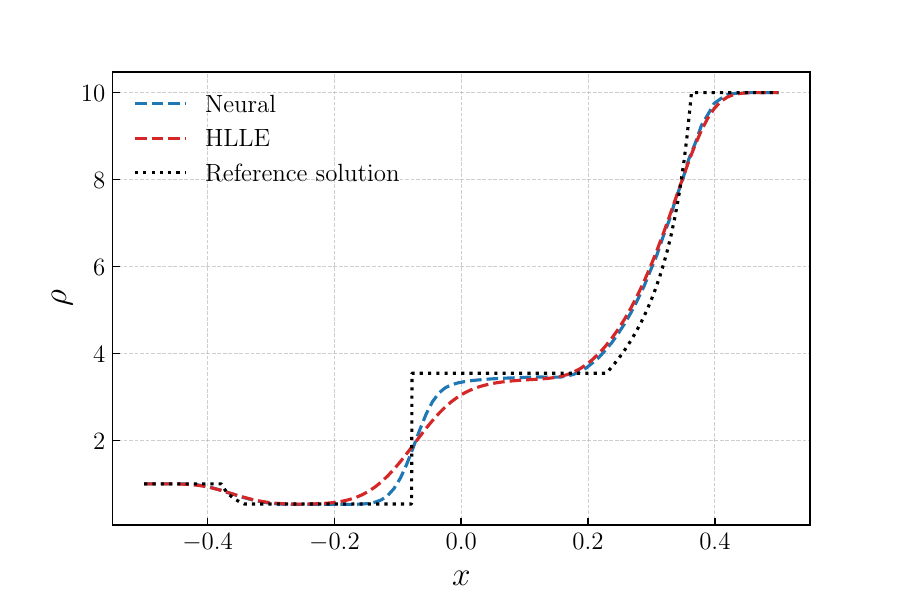}
    \caption{Density profile at $t=0.4$ for the shocktube Problem 1, obtained with the first order scheme on a grid consisting of $100$ points. The exact solution is shown as a dotted line and the numerical results obtained with the HLLE and neural Riemann solvers are shown with dashed lines. While the methods perform similarly at the rarefaction waves, the neural solver resolves the contact discontinuity more accurately than HLLE.}
    \label{fig:shocktube_2_comparison}
\end{figure}

\begin{figure}[htbp]
    \centering
    \includegraphics[width=0.8\textwidth]{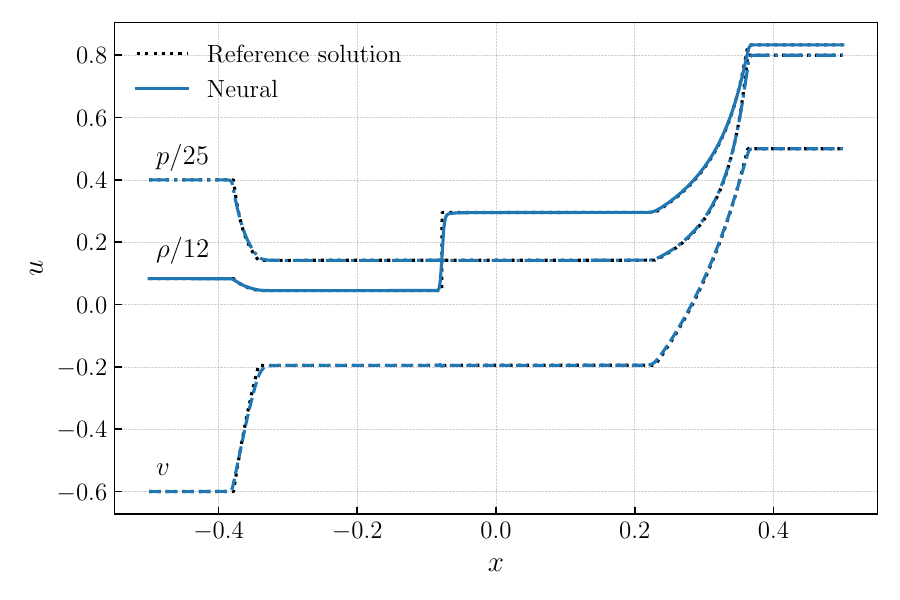}
    \caption{Profiles of density (solid line), pressure (dash-dotted line) and velocity (dashed line) at $t=0.4$ for the shocktube problem 1, obtained with the second order scheme on a grid consisting of $1600$ points. The reference solution is shown as a dotted line.}
    \label{fig:shocktube_2_so}
\end{figure}

As a first test of the solver, we consider a shock tube problem featuring two rarefaction waves propagating in opposite directions. The initial conditions, corresponding to Problem 2 of~\cite{Mignone2005}, are given by: 
\begin{align*} 
    \rho_{\rm L/R} &= 1.0,\, 10.0 \, , \\
    p_{\rm L/R} &= 10.0,\, 20.0 \, , \\
    v_{\rm L/R} &= -0.6,\, 0.5 \, . 
\end{align*} 
This shock tube problem is solved on a domain spanning $x \in [-0.5, 0.5]$, with the initial discontinuity located at $x=0$, using both the first order Godunov scheme as well as the second order WENO-based algorithm.
A comparison of the density profiles obtained with the first order scheme using the neural solver and the traditional HLLE solver on 100 cells is shown in Fig.~\ref{fig:shocktube_2_comparison}.
As can be seen in the figure, both algorithms do comparably well at the rarefactions, but the neural solver outperforms HLLE significantly at the contact discontinuity. The full solution obtained using the neural solver on a grid consisting of $1600$ cells with the second order WENO scheme can be observed in Fig.~\ref{fig:shocktube_2_so}, and $L_1$ error norms on the density obtained with all the Riemann solvers with the first (resp.~second) order scheme at various resolutions are reported in the top left panel of Fig.~\ref{fig:l1_fo} (resp.~Fig.~\ref{fig:l1_so}). Here, the neural Riemann solver can be observed to outperform the traditional counterparts for the first-order at all resolutions, achieving $\sim 25\%$ lower error than HLLC on $1600$ zones. 
This improvement is particularly notable given that both HLLC and the neural solver capture the contact discontinuity with similar sharpness; the superior performance of the neural solver can instead be attributed to its higher accuracy inside the rarefaction waves, highlighting the advantage of explicitly treating distinct wave patterns in the present approach.
While the second-order results are comparable, the HLLC solver results in a $\sim 4\%$ lower error than the neural counterpart. The errors in this setup are found to depend mostly on the reconstruction scheme employed. For instance, when using slope limited reconstruction with either minmod or monotonized-central limiting, the neural solver achieves an error which is lower than HLLC by about $0.5\%$. 
\subsection{Problem 2}
\begin{figure}[htbp]
    \centering
    \includegraphics[width=0.8\textwidth]{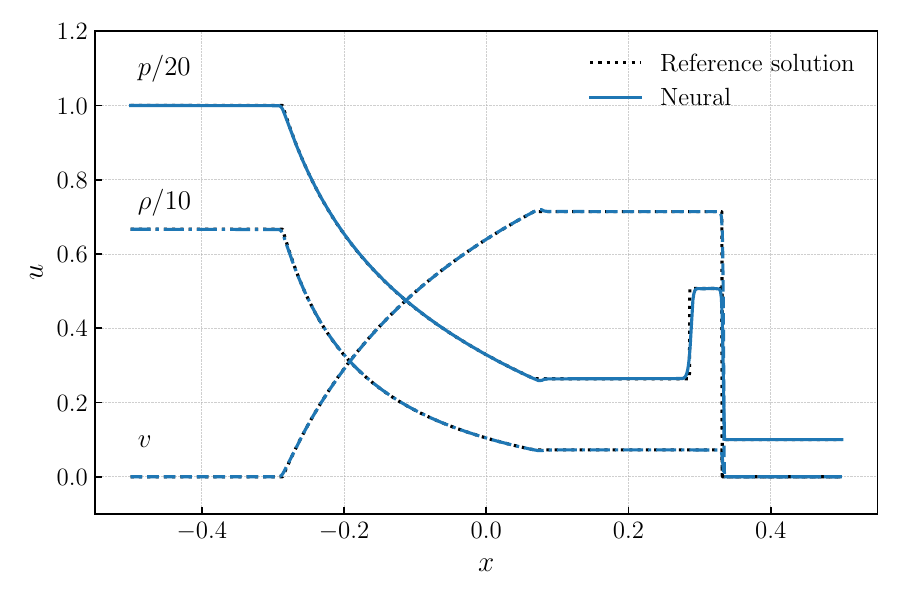}
    \caption{Same as Fig.~\ref{fig:shocktube_2_so} but for Problem 2.}
    \label{fig:shocktube_3_so}
\end{figure}

The second problem we consider is the classic mildly relativistic blastwave benchmark of~\cite{Marti03}. The initial data is given by 
\begin{align*} 
    \rho_{\rm L/R} &= 10.0,\, 1.0 \, , \\
    p_{\rm L/R} &= 40/3,\, 0.0 \, , \\
    v_{\rm L/R} &= 0.0,\, 0.0 \, . 
\end{align*} 
The solution obtained with the second-order WENO scheme and the neural Riemann solver on a grid of $1600$ points is shown in Fig.~\ref{fig:shocktube_3_so}. As can be observed, the solver captures all relevant features of the solution---including the left-going rarefaction, right-moving shock, and contact discontinuity---with excellent accuracy and no signs of instability or spurious oscillations.

The $L_1$ density errors at varying resolutions are reported in the top-right panels of Figs.~\ref{fig:l1_fo} and~\ref{fig:l1_so}, corresponding to the first- and second-order schemes respectively. While in the second-order scheme all solvers perform comparably well, the neural Riemann solver outperforms the HLLE and HLLC solvers in the first-order case, achieving approximately $15\%$ lower error across resolutions.

This test serves as a nontrivial benchmark: the right-moving shock propagates at $v_s = 0.83$, followed by a mildly relativistic contact discontinuity at $v_c = 0.72$. The ability of the learned solver to handle this configuration robustly without any fine-tuning or additional constraints demonstrates its capacity to generalize to physically sharp features within the flow.
\subsection{Problem 3}
\begin{figure}[htbp]
    \centering
    \includegraphics[width=0.8\textwidth]{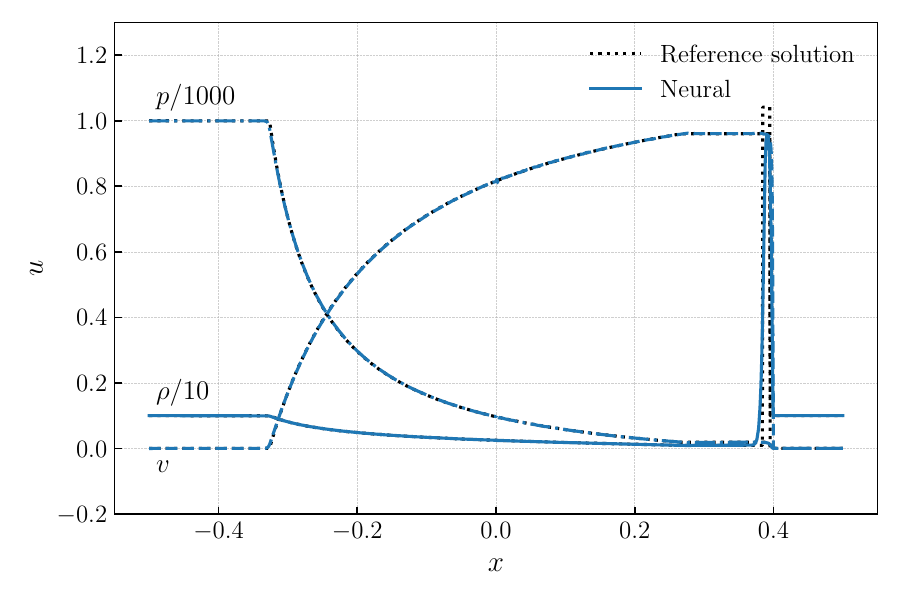}
    \caption{Same as Fig.~\ref{fig:shocktube_2_so} but for Problem 3.}
    \label{fig:shocktube_4_so}
\end{figure}

The next problem is also a relativistic blastwave, except in this case the shock is significantly faster, moving to the right at a speed of $v_s=0.986$. The initial data is given by
\begin{align*} 
    \rho_{\rm L/R} &= 1.0,\, 1.0 \, , \\
    p_{\rm L/R} &= 1000.0,\, 0.01 \, , \\
    v_{\rm L/R} &= 0.0,\, 0.0 \, . 
\end{align*} 
he second-order solution computed with the neural Riemann solver on a grid of $1600$ points is shown in Fig.~\ref{fig:shocktube_4_so}, where the blastwave density is captured to within $\sim 8\%$ of the exact solution. A small ripple is visible near $x=0$ in the solution of the neural solver, which we have found to appear also when the exact counterpart is used. This feature is localized, remains bounded, and converges away with resolution. Its precise origin is unclear, but we note that it does not appear in HLLE or HLLC and does not impact the overall accuracy of the solution.

\begin{figure}[htbp]
    \centering
    \includegraphics[width=0.95\textwidth]{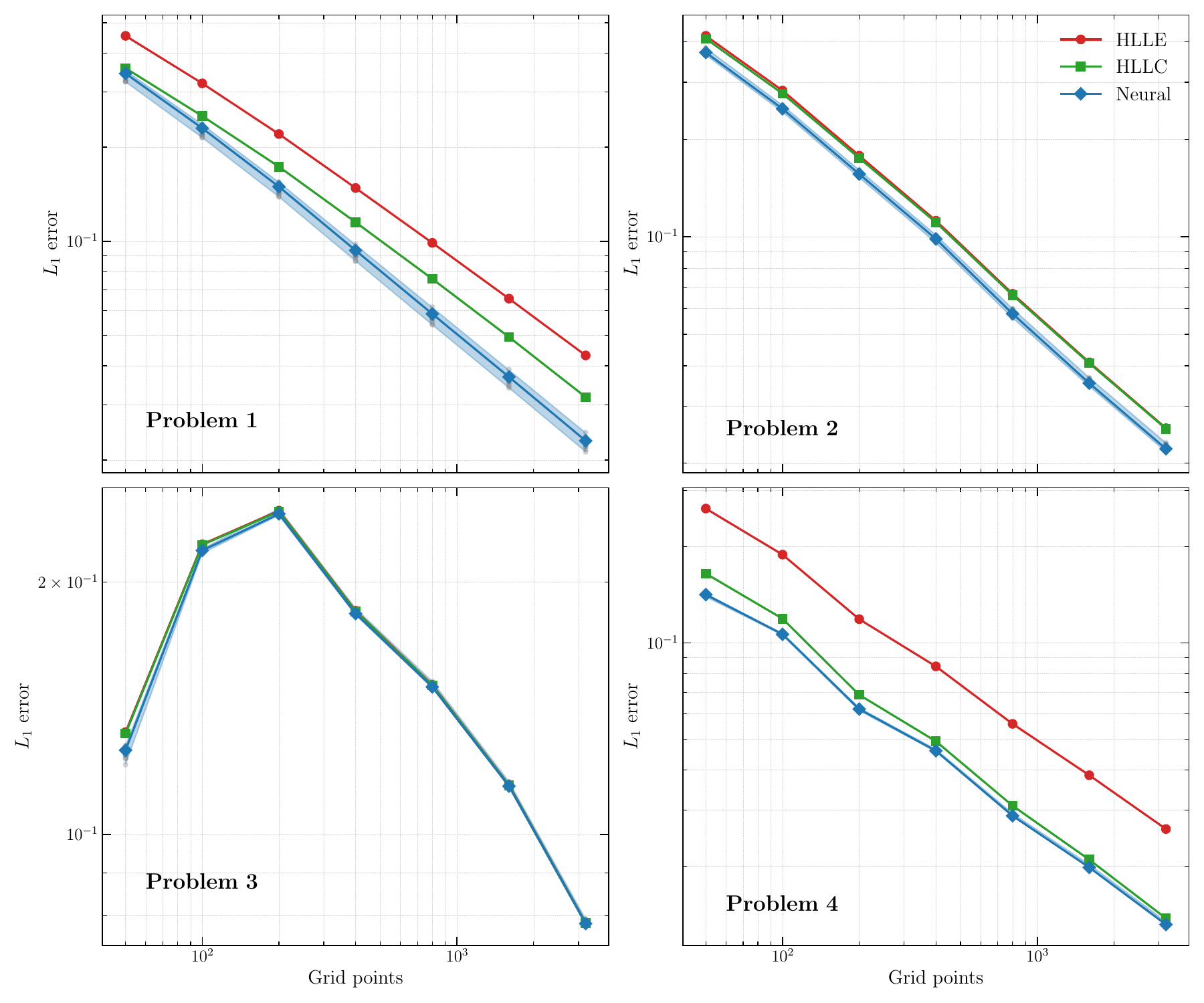}
    \caption{$L_1$ error norms for the density across the four shock tube problems solved using the first-order method, plotted as a function of the number of grid points. The neural solver consistently outperforms HLLE and HLLC across all problems, with the exception of Problem 3, where the blast wave is supersonic.
    Errors from the neural solver corresponding to the main seed (42) are shown as solid lines, while the shaded region represents the range between the 1st and 99th percentiles across all 11 seeds. Individual results from the remaining seeds are overlaid as semi-transparent points.
    A single outlier was excluded from Problem 2 due to a simulation failure.}
    \label{fig:l1_fo}
\end{figure}

The $L_1$ error norms on the rest mass density ($\rho$) are reported in the lower left panels of Figs.~\ref{fig:l1_fo} and~\ref{fig:l1_so}, corresponding to the first- and second-order schemes, respectively. At low resolutions, all solvers significantly overestimate the shock and contact speeds, leading to a non-monotonic error trend, particularly visible in the first-order scheme. This transient behavior is less pronounced in the WENO-based second-order scheme and disappears at higher resolutions.

Beyond this regime, all solvers exhibit comparable convergence behavior. This can be understood by noting that both the shock and contact discontinuity are supersonic, and thus the numerical fluxes become largely insensitive to the internal structure of the Riemann fan. As a result, the differences between the various solvers diminish, and the overall accuracy is dominated by the reconstruction strategy and time integration.
\subsection{Problem 4}
\begin{figure}[htbp]
    \centering
    \includegraphics[width=0.8\textwidth]{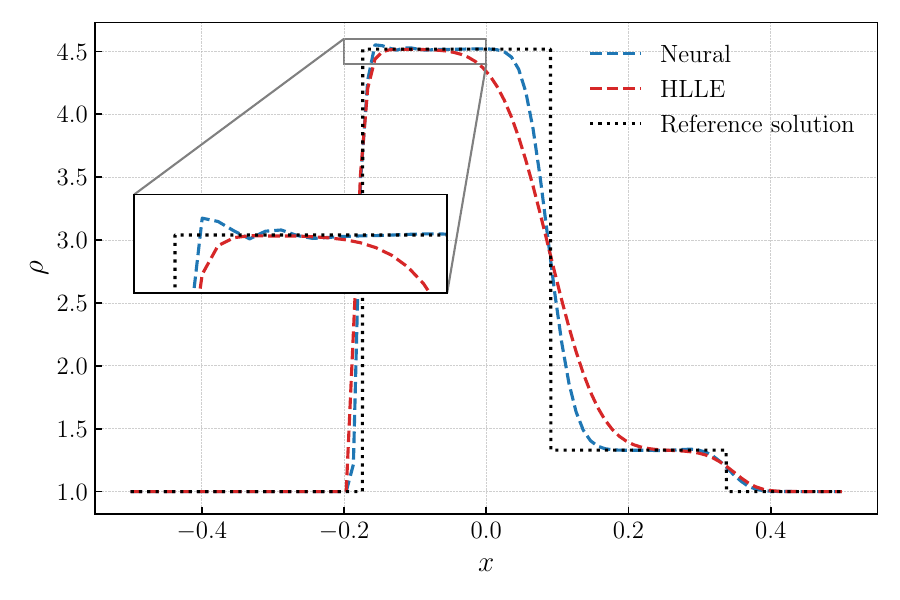}
    \caption{Density profile at $t=0.4$ for the shocktube Problem 4, obtained with the first order scheme on a grid consisting of $100$ points. The exact solution is shown as a dotted line, while the numerical results are shown with dashed lines.}
    \label{fig:shocktube_4_comparison}
\end{figure}

The last problem we consider is a particularly challenging shock tube setup, in which the initial discontinuity decays into two strong shocks propagating in opposite directions. The initial conditions are given by:
\begin{align*} 
    \rho_{\rm L/R} &= 1.0,\, 1.0 \, , \\
    p_{\rm L/R} &= 10.0,\, 1.0 \, , \\
    v_{\rm L/R} &= 0.9,\, 0.0 \, . 
\end{align*} 
Figure~\ref{fig:shocktube_4_comparison} compares the solutions obtained with first-order schemes using the HLLE and neural Riemann solvers. As in previous cases, the neural method exhibits a sharper resolution of the contact discontinuity than its traditional counterpart. However, the zoomed inset reveals that the neural solution also develops small-amplitude oscillations in the post-shock region, even at first order. We note that similar, albeit smaller, oscillations are also visible in the solution obtained using the exact Riemann solver, suggesting that these artifacts are not unique to the neural model but instead reflect the numerical challenges of this strongly shocked configuration.

These oscillations become more pronounced in the second-order scheme. Due to the strength of the shocks in this problem, both the HLLC and neural solvers produce visible post-shock oscillations when combined with higher-order reconstruction. In the case of the neural solver, this is likely attributable to the fact that the steepness of the post-shock profiles amplify the effect of even small errors in the prediction of the contact pressure $p^*$.

\begin{figure}[htbp]
    \centering
    \includegraphics[width=0.8\textwidth]{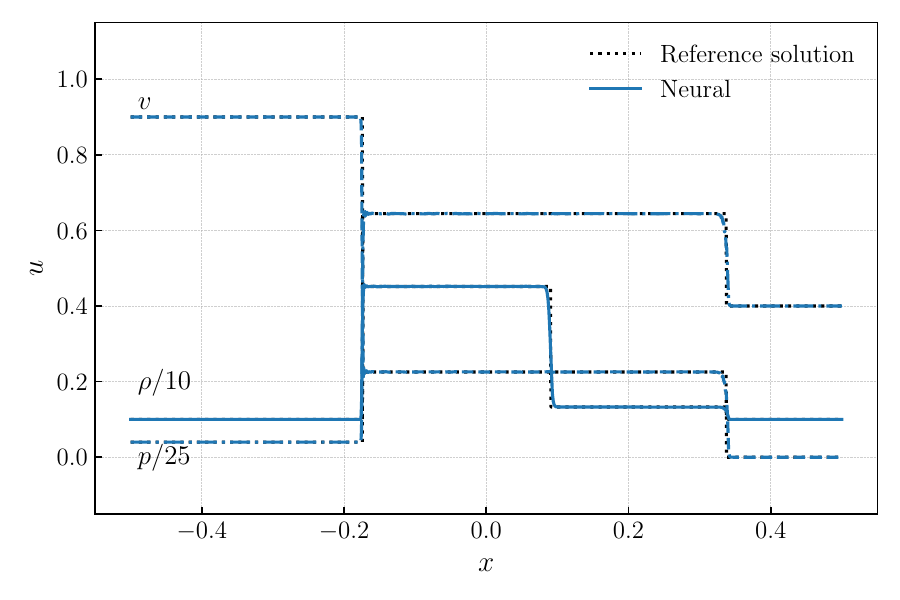}
    \caption{Same as Fig.~\ref{fig:shocktube_2_so} but for Problem 4. The small inconsistency in the predicted contact pressure, together with the sharp high-order reconstruction, produces spurious oscillations in proximity of the shock.}
    \label{fig:shocktube_4_so_1600}
\end{figure}

\begin{figure}[htbp]
    \centering
    \includegraphics[width=0.95\textwidth]{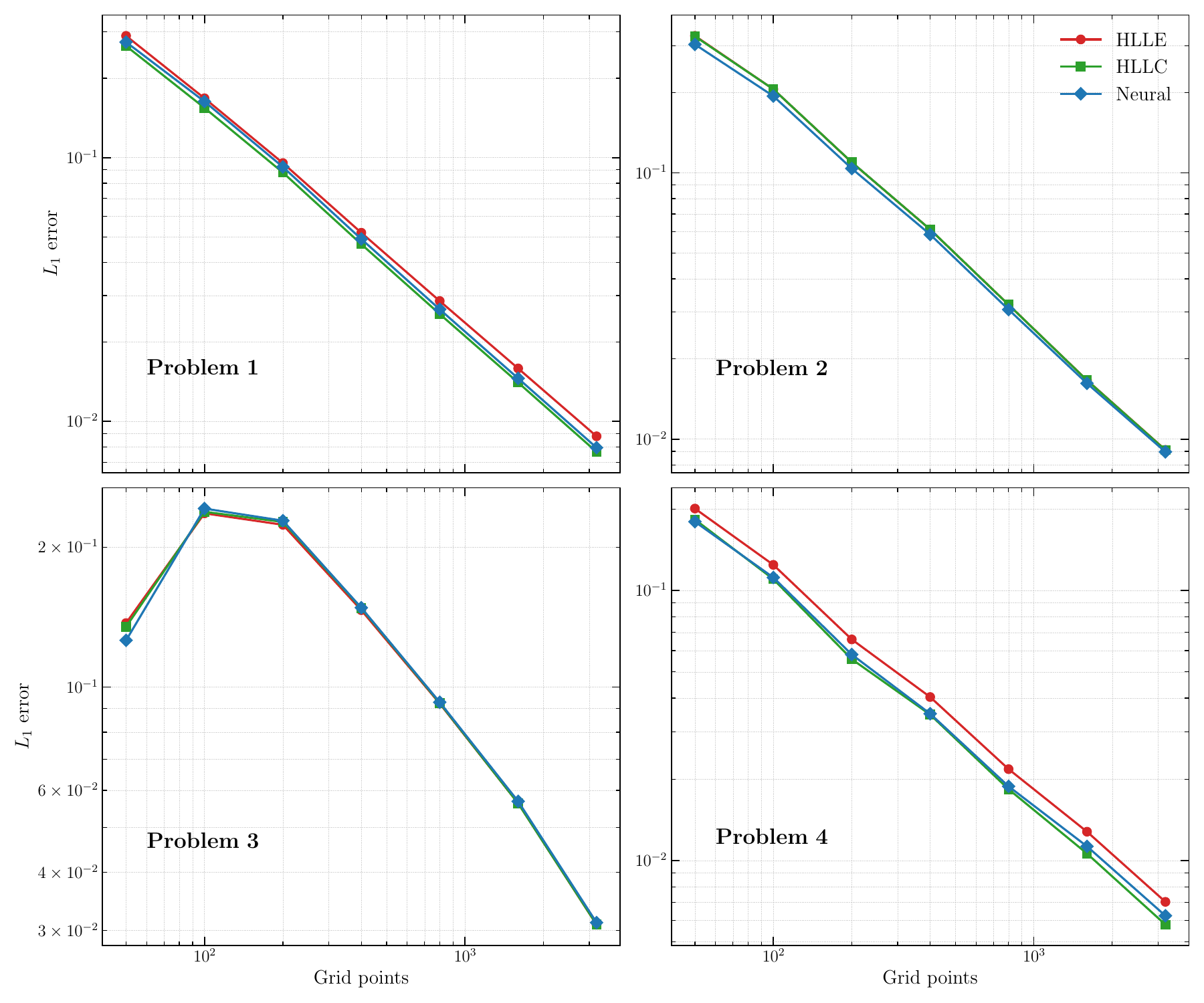}
    \caption{Same as Fig.~\ref{fig:l1_fo}, but for simulations performed with the second-order scheme. Errors are almost identical across solvers, and HLLC outperforms the neural solver in Problem 4, where spurious post-shock oscillations appear. Despite this, the errors are convergent, and the numerical artifacts do not destabilize the solution.}
    \label{fig:l1_so}
\end{figure}

To mitigate this effect, we apply a small amount of flux limiting in the second-order setup, blending in a low-order Rusanov flux computed from cell-averaged states based on the shock indicator proposed in~\cite{Marti96}. While this significantly reduces the oscillations, they remain visible in the pressure and density profiles of the second-order solution at high resolution (see Fig.~\ref{fig:shocktube_4_so_1600}). Despite these artifacts, Fig.~\ref{fig:l1_so} shows that the $L_1$ error continues to converge with increasing resolution, indicating that the oscillations are non-divergent and that the solution remains stable.

\subsection{Performance}
In the previous subsections, we demonstrated that embedding well-constrained machine learning components within a relativistic Riemann solver can lead to improved accuracy compared to traditional approximate solvers. However, the practical viability of this approach hinges not only on its accuracy but also on its computational efficiency. Indeed, the primary reason modern high-resolution shock-capturing codes typically avoid using the exact Riemann solution is its prohibitive cost, especially in the context of three-dimensional simulations in relativistic astrophysics, where millions of Riemann problems must be solved at every timestep.

\begin{figure}[htbp]
    \centering
    \includegraphics[width=0.95\textwidth]{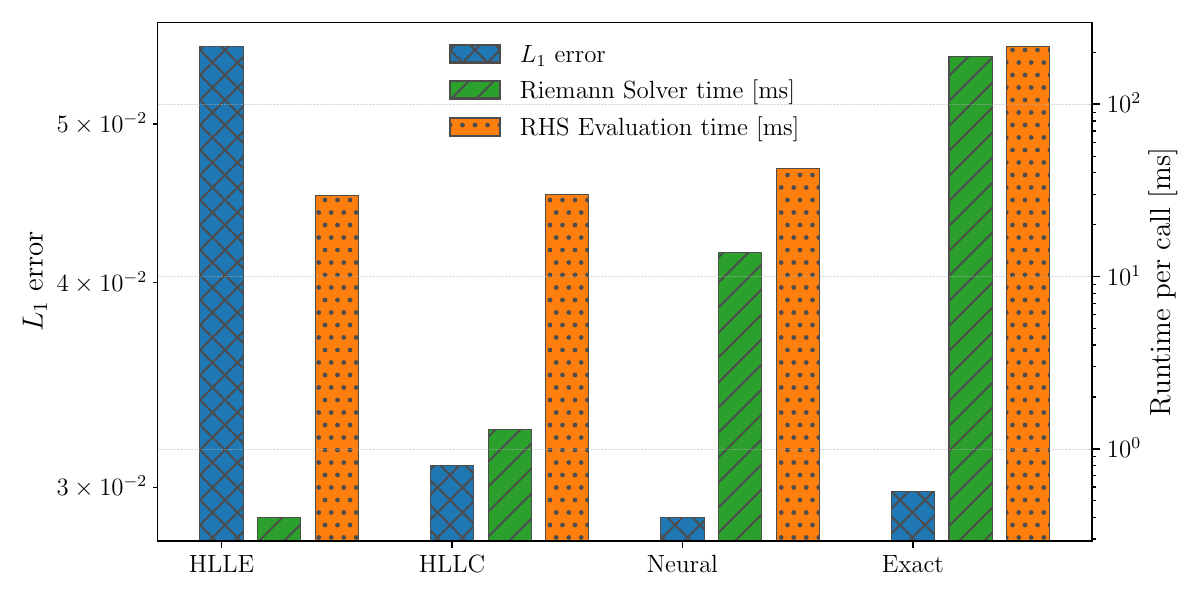}
    \caption{Comparison of $L_1$ density error (blue, left bars), Riemann solver runtime (green, center bars), and total timestep runtime (orange, right bars) for four different Riemann solvers, applied to Problem 4 on a uniform grid with 800 zones using a first-order scheme. All tests were performed on an AMD Mi50 GPU.\@The neural solver matches the accuracy of the exact Riemann solver at significantly reduced computational cost, introducing only a modest increase in total timestep time relative to the HLLC solver.}
    \label{fig:runtime}
\end{figure}

We will now present a runtime analysis of the neural Riemann solver compared to standard solvers. Figure~\ref{fig:runtime} displays a bar chart comparing the $L_1$ density error and runtime for simulations of Problem 4 using a first-order scheme on a uniform grid with 800 zones. The four solvers tested are HLLE, HLLC, the exact solver, and the neural Solver presented in this work.

The results reveal a clear hierarchy in both accuracy and runtime. HLLE is the fastest, which is expected, as HLLE involves only a small number of floating point operations—nine in the central state, neglecting FMA units. However, the blue bars in Figure~\ref{fig:runtime} show that this speed comes at the cost of the largest error among all solvers.

HLLC, while roughly $3.5$ times slower than HLLE, significantly reduces the error, particularly in this test case featuring two strong shocks. Its improved handling of contact discontinuities accounts for this improvement. The exact Riemann solver results in an even lower error, resolving shocks with maximum sharpness—but incurs a computational cost over two orders of magnitude greater than HLLC.

Our neural solver achieves accuracy comparable to the exact solver\footnote{In fact, the error is marginally lower. This discrepancy may stem from occasional failures in the root-finding step (in which case the solver reverts to HLLE), although the precise reason remains unclear.} while being approximately $14$ times faster. Importantly, its runtime is stable and deterministic, unlike the exact solver whose performance varies due to iterative root-finding. This predictability makes the neural solver particularly attractive in the context of large-scale, highly parallel simulations. 

It is also worth noting that despite the stark differences in per-interface solver cost, the total simulation runtime is not dominated by the Riemann solver. As shown by the orange bars in Figure~\ref{fig:runtime}, which represent the full timestep wall time for each configuration, using the neural solver increases the overall timestep cost by about a factor of $1.5$ compared to HLLC.

Finally, it is worth noting that the structure of the neural Riemann solver---being composed of compact, feedforward neural networks---makes it particularly well suited to GPU execution. While traditional approximate solvers already approach peak throughput for a small number of floating-point operations, and exact solvers suffer from irregular control flow and root-finding overheads, neural network inference is highly parallelizable and can benefit substantially from advances in GPU hardware and compiler technology. As relativistic hydrodynamics codes increasingly adopt GPU acceleration~\cite{Stone2024,Kalinani2024,Shankar2023}, neural solvers may therefore become an even more attractive alternative from a performance standpoint.
\FloatBarrier
\section{Future Directions}~\label{sec:future}
The results presented in this work demonstrate the potential of machine learning techniques to enhance the performance of traditional Riemann solvers in relativistic hydrodynamics. However, several avenues for future research remain open.

Firstly, the present work is restricted to one spatial dimension. In most HRSC codes, the Riemann problem remains one-dimensional due to dimensional splitting of the fluxes. However, the presence of non-zero transverse velocity components complicates the solution due to the relativistic coupling of velocity components in the Lorentz factor.

Although this introduces some theoretical and numerical challenges, it does not fundamentally alter the structure of the Riemann problem. Indeed, exact solutions exist for multi-dimensional flows~\cite{Pons00}. It appears that---despite the increased complexity in training data generation---an approach similar to the one presented here, augmented with appropriately enhanced networks, could be extended to multi-dimensional applications while retaining its accuracy, interpretability, and computational efficiency.

Secondly, while the method in its present form effectively demonstrates that the collection of networks trained on a problem-agnostic dataset can reproduce exact solutions to a high degree of accuracy in a variety of regimes, the sharpness of the resulting fluxes and the inherent inaccuracy of the network prediction can lead to numerical artifacts in the solution. This is especially true for high-order schemes, where---analogously to what happens for traditional Riemann solvers---the choice of reconstruction can significantly impact the quality of the resulting solution. Evidently, the application of this method to state-of-the-art simulations will require further work to improve its robustness. Several promising avenues for this include:
\begin{itemize}
    \item Using the target function residual as an inexpensive error estimate, with a fallback to a more diffusive solver (e.g., HLLE) if the residual exceeds a predefined threshold.
    \item Applying a more aggressive flux limiting strategy in the vicinity of strong shocks, where small errors in the network prediction are more likely to lead to spurious numerical effects in the solution.
    \item Implementing a more advanced data sampling algorithm, potentially including a curriculum learning approach where the networks are gradually exposed to increasingly difficult regimes.
\end{itemize}

\noindent A complementary approach to improve solver robustness, based on network ensembling and residual-based selection, is presented in Appendix~\ref{appendix:ensemble}.

Additionally, we note that the learned components of the neural solver implicitly encode EOS-specific information through the training data. While the present study is restricted to a gamma-law EOS, generalizing to more complex or tabulated EOSs would require tailored data generation and potentially additional network components to capture the increased complexity.

Finally, although the tests considered here span multiple wave patterns, relativistic regimes, and orders of magnitude in physical quantities, they still represent a subset of the configurations relevant to astrophysical and heavy-ion collision applications. Capturing such extreme scenarios may require larger networks trained on appropriately scaled datasets, or a modular strategy involving specialized sub-networks, each targeting a specific region of the input space.

\FloatBarrier
\section{Conclusions}\label{sec:conclusions}
We have introduced a data-driven framework for solving the Riemann problem in relativistic hydrodynamics, replacing online root-finding with neural networks trained on physically constrained targets. By decomposing the problem into distinct wave pattern categories and training specialized networks for each, we developed a hybrid solver that achieves accuracy comparable to the exact solution, while significantly reducing computational cost.

A key strength of this approach lies in its modular design: the neural components are compact, well-scoped, and embedded within a traditional physics-based solver. Rather than replacing the full flux computation with a black-box model, we substitute only the computational bottlenecks with learned components, allowing the overall method to remain interpretable, robust, and closely aligned with the underlying physics.

Through a series of canonical shock tube tests, we demonstrated that the same trained networks perform reliably across a broad range of physical regimes, including problems featuring strong shocks and contact discontinuities. Compared to traditional approximate solvers such as HLLE and HLLC, the neural solver offers improved accuracy while maintaining robustness and avoiding the runtime overhead of root-finding. Its non-iterative structure ensures predictable performance and makes it particularly well-suited for execution on modern GPU architectures, positioning it as a promising candidate for large-scale, high-resolution simulations.

Despite these promising results, we find that the inherent approximation and non-determinism of neural network inference can lead to small numerical artifacts in especially challenging regions of the flow. However, our tests show that these remain bounded and do not spoil convergence of the overall scheme. Future improvements could involve uncertainty-aware inference (e.g., ensembling or residual-based fallback) or adaptive flux limiting to mitigate such artifacts more systematically.

While this study focuses on the one-dimensional case, the framework is general and extensible to higher dimensions. The main challenge lies in generating representative training data and potentially extending the learned components to account for effects that are analytic in one spatial dimension but become more complex in higher dimensions, rather than modifying the architecture itself. Analogously, while we restrict our attention to a simple gamma-law equation of state, extending the method to more complex or tabulated EOSs would require similar adjustments in training and model design. These generalizations, along with applications to more complex systems such as relativistic magnetohydrodynamics, will be explored in future work.

Overall, the results suggest that machine-learned Riemann solvers, trained once and deployed widely, offer a compelling approach to achieving high efficiency and physical fidelity across a wide range of relativistic hydrodynamic applications. Additional strategies such as model ensembling, discussed in Appendix~\ref{appendix:ensemble}, further enhance robustness and may be key to future deployment in production-scale simulations.

\section{Data and Code Availability}
The code used to generate all results in this paper is openly available at our \href{https://github.com/carlomusolino/neural_riemann_solver}{GitHub repository}. Trained model weights and datasets are available from the author upon reasonable request.

\section*{Acknowledgements}
It is a pleasure to thank Prof.~Dr.~L.~Rezzolla, Dr.~T.~Gorda and Prof.~Dr.~R.~De Pietri for their careful reading of the manuscript and for their suggestions, which contributed to improving the quality of this work.
We am also sincerely grateful to Dr.~C.~Ecker and Dr.~F.~Camilloni, for insightful discussions during the development of this method.
All calculations were performed on the GPU partition of the supercomputer at the Institute for Theoretical Physics, Goethe University Frankfurt.

\paragraph{Funding information}
This work was funded by the ERC Advanced Grant ``JETSET: Launching, propagation and emission of relativistic jets from binary mergers and across mass scales'' (Grant No.~884631).

\begin{appendix}

\section{Data generation algorithms}
\label{appendix:data_generation}

As discussed in the main text, all the learned components of the neural Riemann solvers---namely, the pressure predictor networks $g_p$ and the rarefaction solvers $r_\leftarrow$ and $r_\rightarrow$---are simple feedforward MLPs mapping multi-dimensional inputs to scalar outputs. All these networks serve the same functional purpose: to find the root $x^*$ of a nonlinear function $f(x, \theta)$ given parameters $\theta$ (refer to Eqs.~\eqref{eq:solution_map} and~\eqref{eq:rootfinding_general} for context). Training is conducted using a composite loss that includes (i) the difference between the predicted and true roots and (ii) the residual of the function at the network-predicted solution.

Although the data generation strategy follows similar steps for each network, key differences arise in how the relevant region of the input space is sampled  for each solver. Throughout this section, we assume without loss of generality that $p_{\rm L} \geq p_{\rm R}$.

\vspace{1em}
\noindent
\paragraph*{Pressure predictors \boldmath{$g_p$} ($p \in \{{\rm 2S}, {\rm 2R}, {\rm SR}\}$)}
\begin{enumerate}
    \item Generate $N$ samples $\{\theta_i\}_{i=1,\ldots,N}$ in $[0,1]^6$ using a quasi-random Sobol engine.
    \item Rescale the first five components of each $\theta_i$ to represent $\log \rho_{\rm L}$, $\log \rho_{\rm R}$, $\log p_{\rm R}$, $\log (p_{\rm L}/p_{\rm R})$, $v_{\rm L}$. The sixth component is left unscaled initially.
    \item For each rescaled $\theta_i$, compute the pattern-specific limiting relative velocities $\Delta v_p^{\rm min}, \, \Delta v_p^{\rm max}$.
    \item Rescale the sixth component of $\theta_i$ to span the allowed invariant relative velocity range $\Delta v_{\rm LR} \in [\Delta v_p^{\rm min}, \Delta v_p^{\rm max}]$.
    \item Compute $v_{\rm R}$ from $v_{\rm L}$ and $\Delta v_{\rm LR}$.
    \item Solve the corresponding nonlinear equation to obtain the true contact pressure $p^*$ for each input.
\end{enumerate}

\noindent This procedure defines the left and right states while indirectly controlling the relative velocity, which is constrained to lie within the range compatible with the chosen wave pattern.

\vspace{1em}
\noindent
\paragraph*{Rarefaction solvers \boldmath{$r_\leftarrow$}, \boldmath{$r_\rightarrow$}}
\begin{enumerate}
    \item Generate $N$ samples $\{\theta_i\}_{i=1,\ldots,N}$ in $[0,1]^3$ using a Sobol engine.
    \item Rescale each $\theta_i$ to represent $(\log \rho_a, \log p_a, v_a)$, where $a$ denotes the post-rarefaction state.
    \item Exclude supersonic states where $c_s(p_a) < \pm v_a$—with the plus (minus) sign corresponding to the right-going (left-going) wave—as such states cannot constitute the head of a rarefaction.
    \item Exclude states for which the rarefaction fails to intersect the interface at $\xi = 0$. Since $\xi(p)$ is a monotonic function, it is sufficient to evaluate $\xi_0 = \xi(p=0)$ via Eqs.~\eqref{eq:jpm} and~\eqref{eq:cs_xi} and exclude states for which $\xi_0 < 0$ (left-going) or $\xi_0 > 0$ (right-going).
    \item For the remaining inputs, solve the corresponding nonlinear equation to obtain the sound speed at $\xi = 0$.
\end{enumerate}

For all networks, the training dataset spans densities in $[10^{-2}, 10^2]$, pressures in $[10^{-7}, 10^{3.5}]$, and velocities in $[-0.99, 0.99]$ (in units where $c = 1$). The range of pressure jumps $\log_{10}(p_{\rm L}/p_{\rm R})$ varies by solver: $[0.1, 1.5]$ for the double shock solver, $[0.1, 1.0]$ for the double rarefaction solver, and $[0.1, 7.0]$ for the shock-rarefaction solver.

All training and inference operations are carried out using double precision in 
\texttt{PyTorch} to maintain consistency with the numerical solver and reduce round-off sensitivity.

\section{Model ablation study}\label{appendix:ablation}
\subsection{Dependence on random seed} \label{appendix:training}
\begin{figure}[htbp]
    \centering

    % First row
    \begin{subfigure}[b]{0.3\textwidth}
        \centering
        \includegraphics[width=\textwidth]{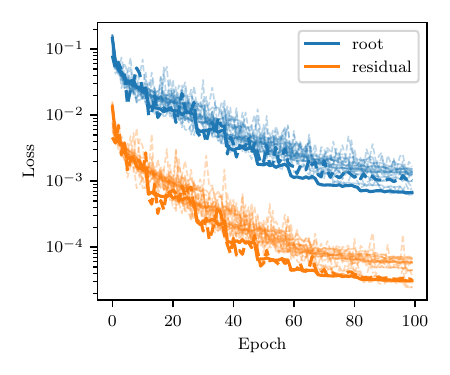}
        \caption{$g_{\rm 2S}$}
    \end{subfigure}
    \hfill
    \begin{subfigure}[b]{0.3\textwidth}
        \centering
        \includegraphics[width=\textwidth]{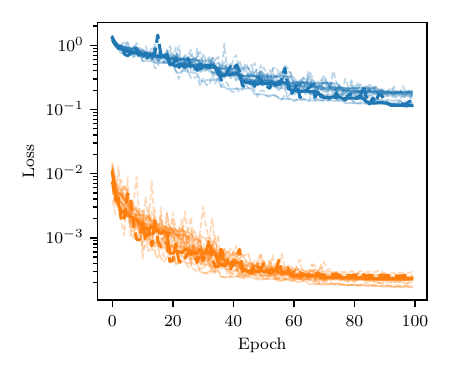}
        \caption{$g_{\rm 2R}$}
    \end{subfigure}
    \hfill
    \begin{subfigure}[b]{0.3\textwidth}
        \centering
        \includegraphics[width=\textwidth]{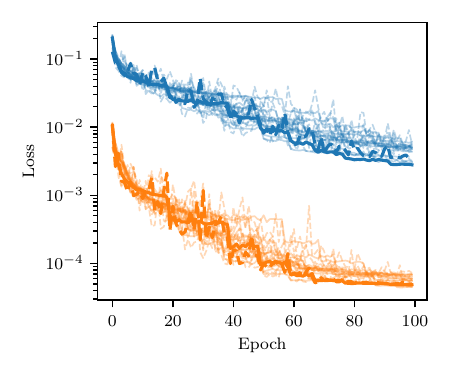}
        \caption{$g_{\rm SR}$}
    \end{subfigure}

    \vspace{1em}

    % Second row: centered with horizontal spacing
    \hspace*{0.15\textwidth} % left padding
    \begin{subfigure}[b]{0.3\textwidth}
        \centering
        \includegraphics[width=\textwidth]{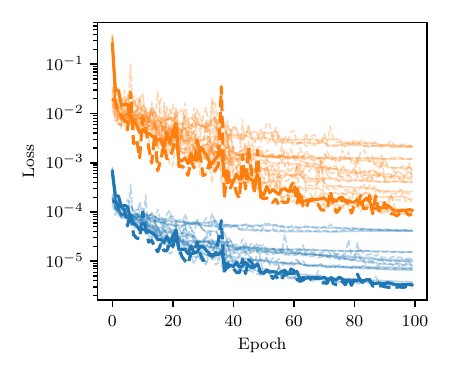}
        \caption{$r_\leftarrow$}
    \end{subfigure}
    \hfill % gap between the two figs
    \begin{subfigure}[b]{0.3\textwidth}
        \centering
        \includegraphics[width=\textwidth]{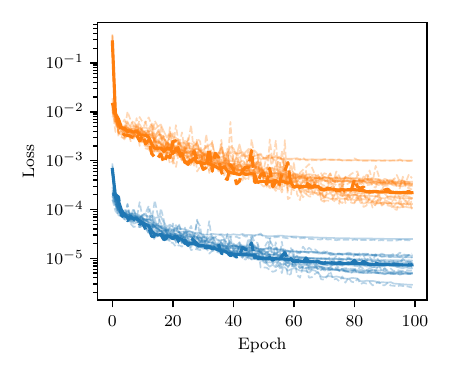}
        \caption{$r_\rightarrow$}
    \end{subfigure}
    \hspace*{0.15\textwidth} % left padding

    \caption{Training and test loss trajectories (root in blue and residual in orange) for each learned Riemann solver component. Solid lines indicate training loss; dashed lines indicate test loss. Main seed (42) shown in full opacity; others shown more faintly to indicate variance across seeds.}
    \label{fig:loss_trajectories}
\end{figure}

Loss curves for $11$ different seeds (i.e.~varying initial weights, Sobol sequences, and batch orderings) where the model and dataset sizes are kept fixed to the ones used in the main text are shown in the panels of Fig.~\ref{fig:loss_trajectories}. As shown, training and test losses follow similar trajectories for all solvers, indicating a low risk of overfitting. The variability across seeds is minimal, although slightly more pronounced in the rarefaction solvers. Moreover, the double rarefaction solver appears to converge to relatively large root errors (on the order of $10^{-1}$) despite low residuals, suggesting that the training signal may be weak in regions close to vacuum, where the residual function flattens and gradients vanish.  
\subsection{Dependence on dataset size}\label{appendix:dataset_size}
\begin{figure}[htbp]
    \centering

    % First row
    \begin{subfigure}[b]{0.3\textwidth}
        \centering
        \includegraphics[width=\textwidth]{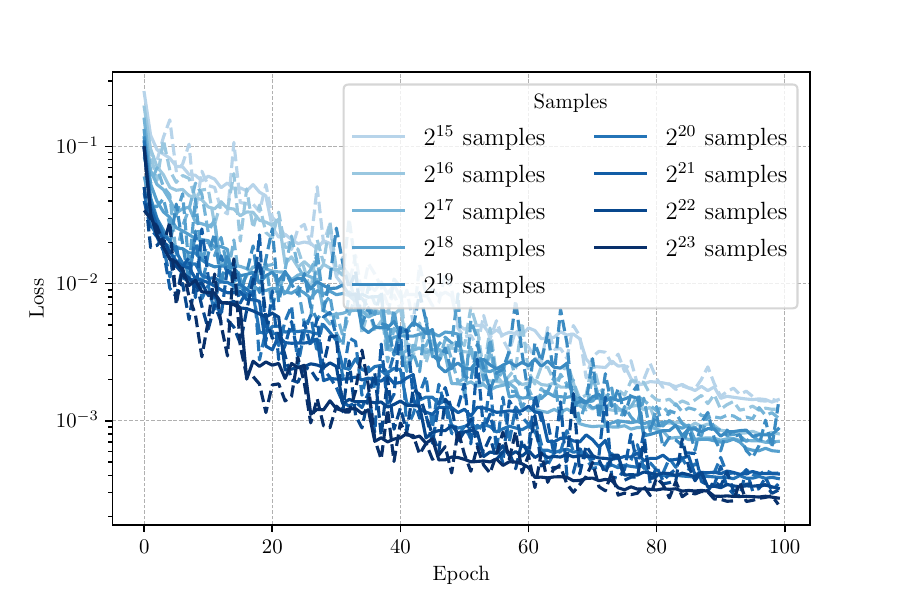}
        \caption{$g_{\rm 2S}$}
    \end{subfigure}
    \hfill
    \begin{subfigure}[b]{0.3\textwidth}
        \centering
        \includegraphics[width=\textwidth]{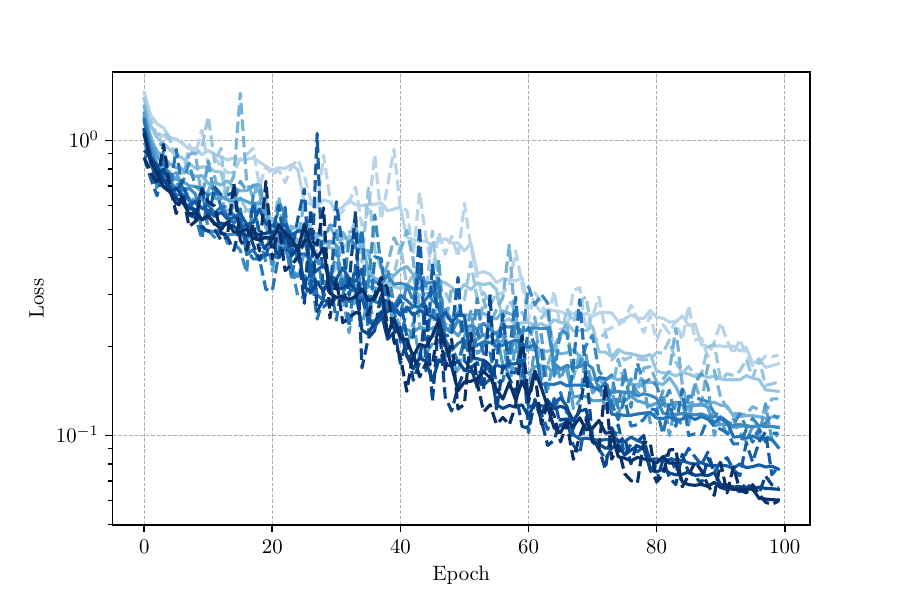}
        \caption{$g_{\rm 2R}$}
    \end{subfigure}
    \hfill
    \begin{subfigure}[b]{0.3\textwidth}
        \centering
        \includegraphics[width=\textwidth]{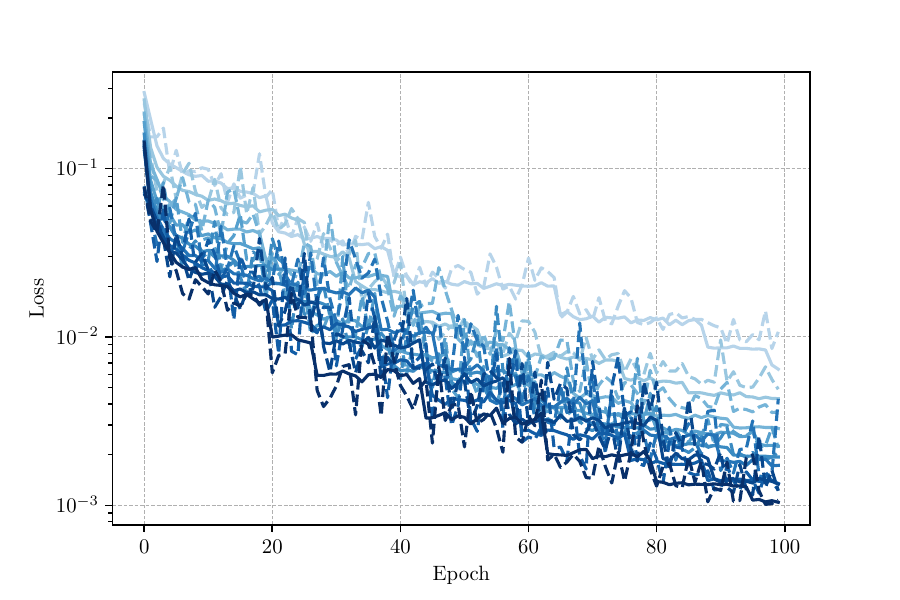}
        \caption{$g_{\rm SR}$}
    \end{subfigure}

    \vspace{1em}

    % Second row: centered with horizontal spacing
    \hspace*{0.15\textwidth} % left padding
    \begin{subfigure}[b]{0.3\textwidth}
        \centering
        \includegraphics[width=\textwidth]{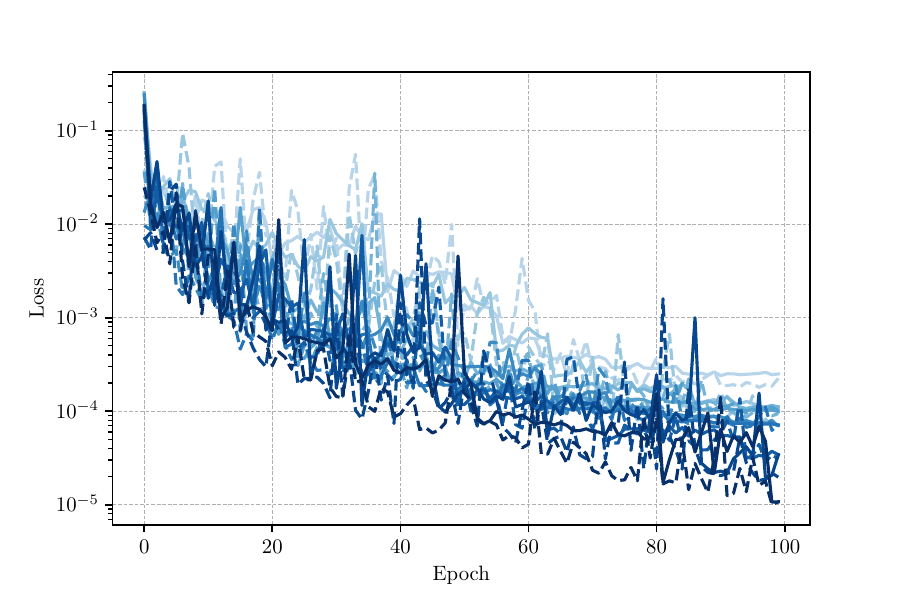}
        \caption{$r_\leftarrow$}
    \end{subfigure}
    \hfill % gap between the two figs
    \begin{subfigure}[b]{0.3\textwidth}
        \centering
        \includegraphics[width=\textwidth]{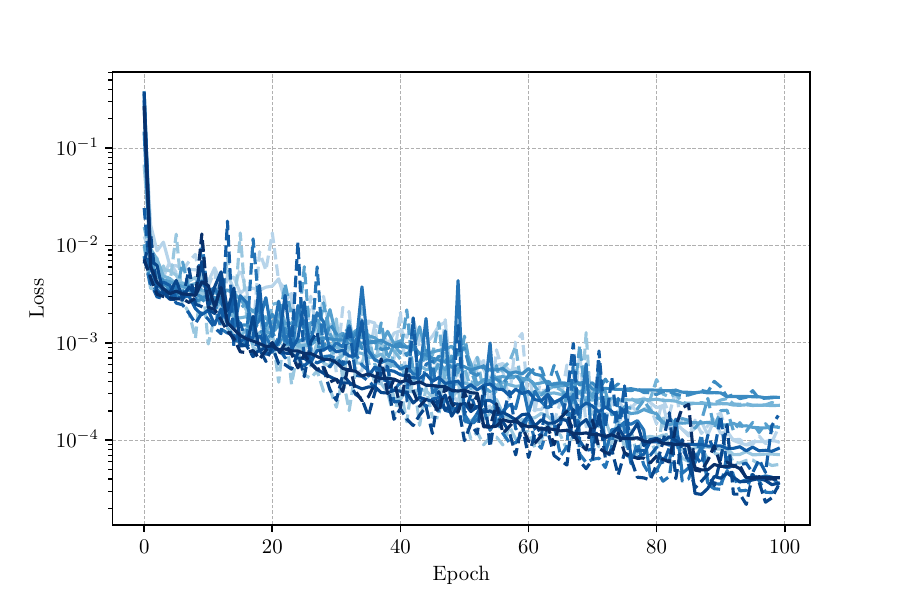}
        \caption{$r_\rightarrow$}
    \end{subfigure}
    \hspace*{0.15\textwidth} % left padding

    \caption{Training and test loss trajectories for each learned Riemann solver component with various dataset sizes. Solid lines indicate training loss; dashed lines indicate test loss.}
    \label{fig:loss_trajectories_dset}
\end{figure}

Figure~\ref{fig:loss_trajectories_dset} shows training curves for all components of the neural solver trained on dataset sizes ranging from $2^{14}$ ($\sim 1.6\times 10^4$) to $2^{23}$ ($\sim 8\times 10^6$). To keep the overall training effort consistent, we scale the batch size proportionally to the dataset size: the smallest dataset ($2^{14}$ samples) uses a batch size of $16$, while the largest ($2^{23}$) uses $8192$. This ensures that the total number of weight updates remains approximately constant across all runs.

As shown in the figure, both training and validation losses generally improve with dataset size, although the trend is not strictly monotonic. In particular, the rarefaction solvers occasionally plateau at intermediate dataset sizes, suggesting that some regions of the parameter space are either sparsely sampled or especially difficult to learn.

\begin{figure}[htbp]
    \centering

    % First row
    \begin{subfigure}[b]{0.45\textwidth}
        \centering
        \includegraphics[width=\textwidth]{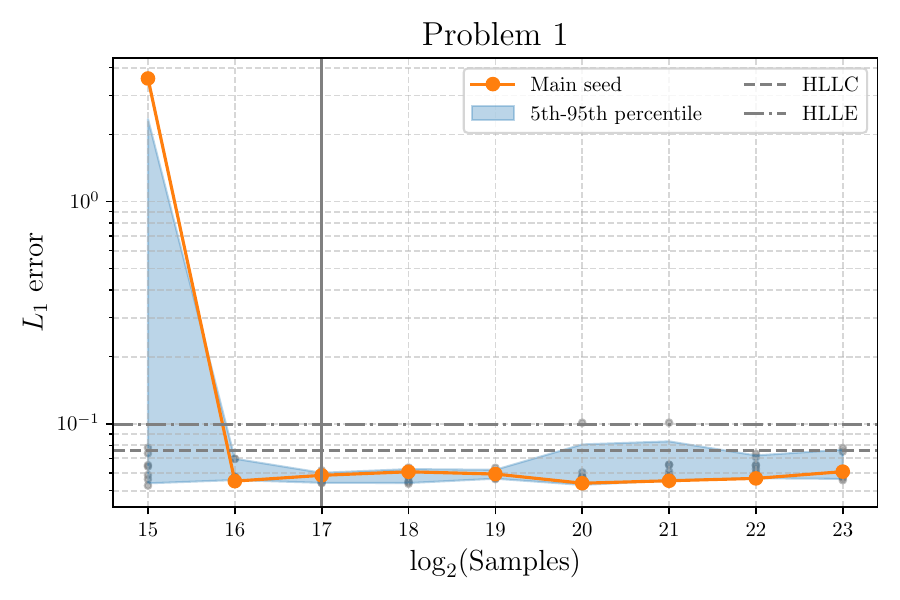}

    \end{subfigure}
    \hfill
    \begin{subfigure}[b]{0.45\textwidth}
        \centering
        \includegraphics[width=\textwidth]{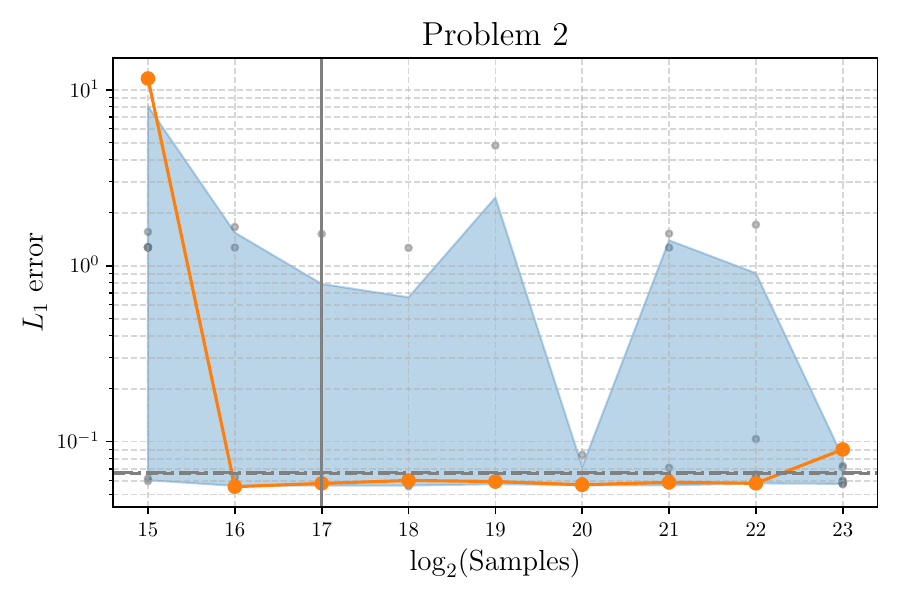}
    \end{subfigure}

    \vspace{1em}

    % Second row: centered with horizontal spacing

    \begin{subfigure}[b]{0.45\textwidth}
        \centering
        \includegraphics[width=\textwidth]{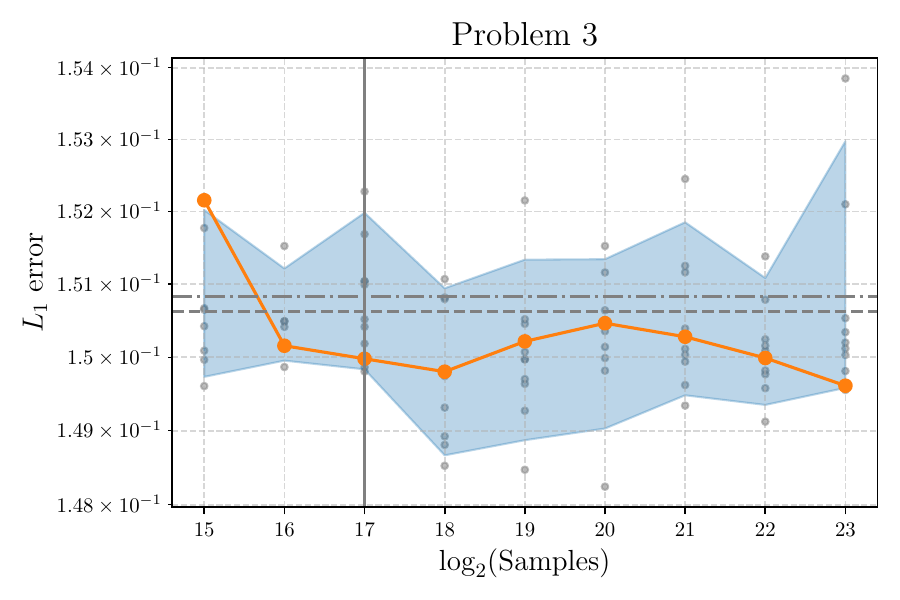}
    \end{subfigure}
    \hfill % gap between the two figs
    \begin{subfigure}[b]{0.45\textwidth}
        \centering
        \includegraphics[width=\textwidth]{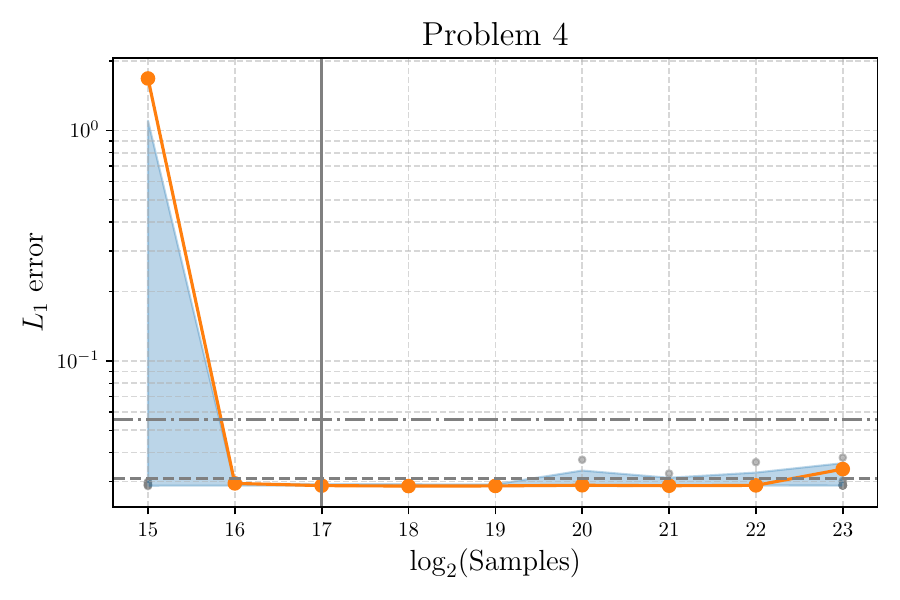}
    \end{subfigure}

    \caption{$L_1$ density error across problems for networks with two hidden layers of 64 neurons each. The rarefaction networks have half as many neurons as the pressure predictors and are trained on a dataset $1/10$ the size. Seed-to-seed variability is shown as a shaded region, while the main seed (42) is shown with full opacity. Across most dataset sizes, the errors exhibit relatively little variation across seeds, except for very small or very large datasets. The only major exception is Problem 2, where a small number of outlier seeds consistently appear. This behavior likely stems from the fact that Problem 2 features an extreme pressure jump (of order $10^7$) in the initial data, placing it near the boundary of the synthetic data distribution.
    The vertical solid line marks the dataset size used in the main text ($2^{17}$ samples), which consistently shows minimal seed-to-seed variability across all problems.}
    \label{fig:l1_dset}
\end{figure}

From a practical standpoint, the most important question is how these effects propagate to simulation accuracy. Figure~\ref{fig:l1_dset} reports the $L_1$ density error across the four canonical shock tube problems introduced in the main text. All simulations are performed using a first-order scheme on a uniform grid of $800$ zones, and the $L_1$ errors are computed against the exact solution at $t=0.4$.

Each panel shows the performance of the main seed (42) as a solid line, with the $5^{\rm th}$ to $95^{\rm th}$ percentile range across all seeds shown as a shaded region. The vertical dashed line marks the dataset size ($2^{17}$ samples) used in the main text, where seed-to-seed variability is consistently minimized across all problems.

Overall, the neural solver appears to be robust across a broad range of dataset sizes. For Problems 1 and 4, all seeds perform better than HLLC in the range $16 \leq \log_2(N) \leq 19$, with only marginal degradation at larger dataset sizes. Problem 3 shows moderate variability independent of dataset size, with the best and worst seeds differing by no more than $\sim 2\%$.

The main exception is Problem 2, where a small number of outlier seeds consistently appear where the simulation fails completely, and the error exceeds that of the approximate solver by more than one order of magnitude. This behavior likely stems from the fact that Problem 2 features an extreme pressure jump (of order $10^7$) in the initial conditions, placing it near the boundary of the synthetic data distribution. Addressing this issue may require improvements in dataset generation, or runtime error recovery strategies such as those outlined in Section~\ref{sec:future}.

Importantly, these results highlight that the training loss is not always a reliable proxy for real-world solver performance. All loss curves in Figure~\ref{fig:loss_trajectories_dset} show smooth, stable convergence to low loss values, even for seeds that later exhibit degraded simulation accuracy. This mismatch is especially pronounced at larger dataset sizes, where the loss may be dominated by easier samples and miss localized failure modes.

Nevertheless, the fact that most seeds succeed across all problems—and that a conservative dataset size can largely suppress seed-to-seed variability—demonstrates the potential of the method. We view these results as a promising first step toward a robust, generalizable neural Riemann solver.

\subsection{Dependence on model size}
\label{appendix:model_size}
\begin{figure}[htbp]
    \centering

    % First row
    \begin{subfigure}[b]{0.45\textwidth}
        \centering
        \includegraphics[width=\textwidth]{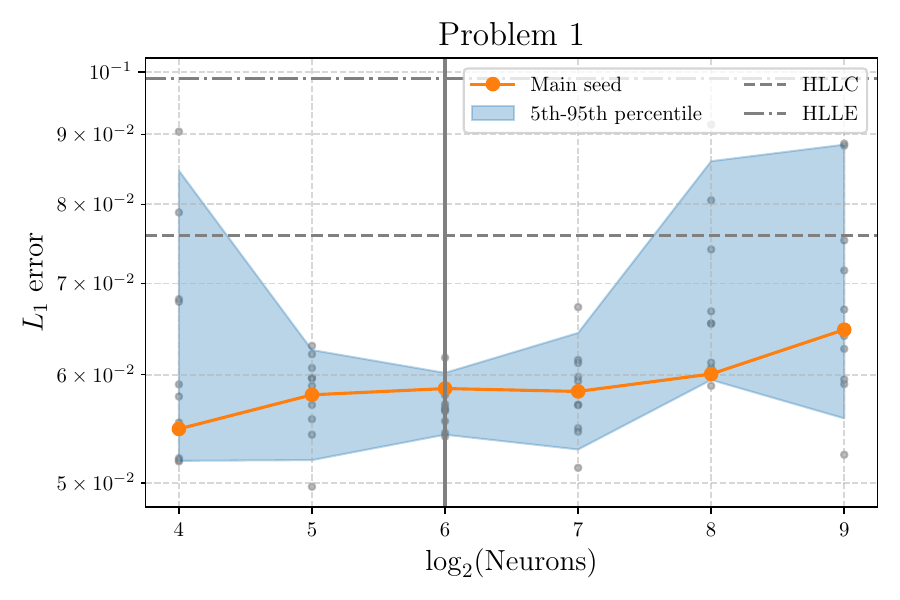}

    \end{subfigure}
    \hfill
    \begin{subfigure}[b]{0.45\textwidth}
        \centering
        \includegraphics[width=\textwidth]{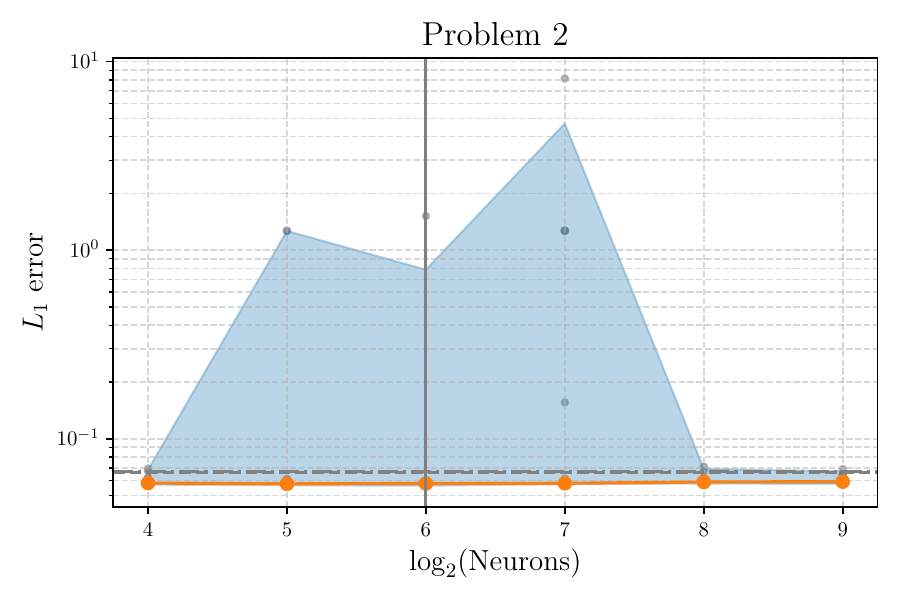}
    \end{subfigure}

    \vspace{1em}

    % Second row: centered with horizontal spacing

    \begin{subfigure}[b]{0.45\textwidth}
        \centering
        \includegraphics[width=\textwidth]{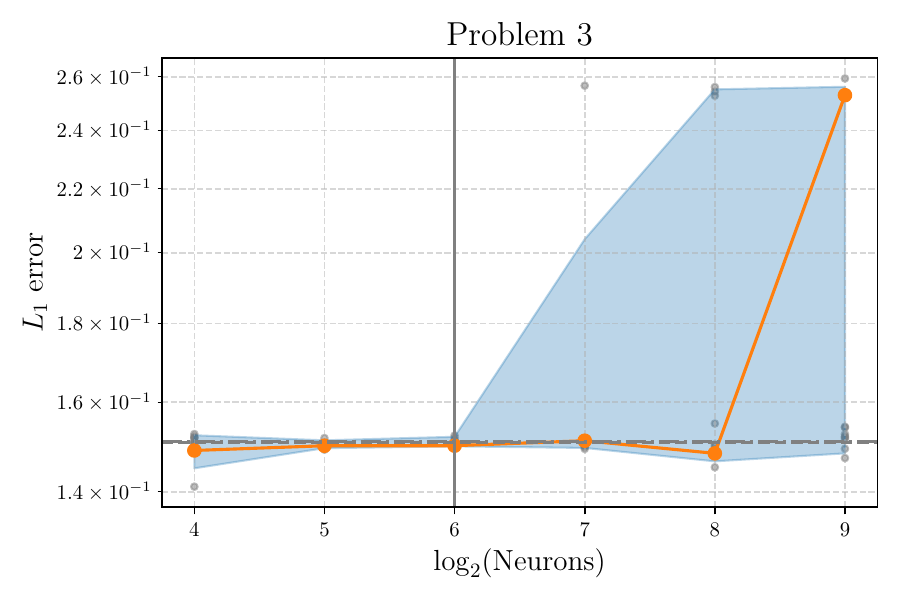}
    \end{subfigure}
    \hfill % gap between the two figs
    \begin{subfigure}[b]{0.45\textwidth}
        \centering
        \includegraphics[width=\textwidth]{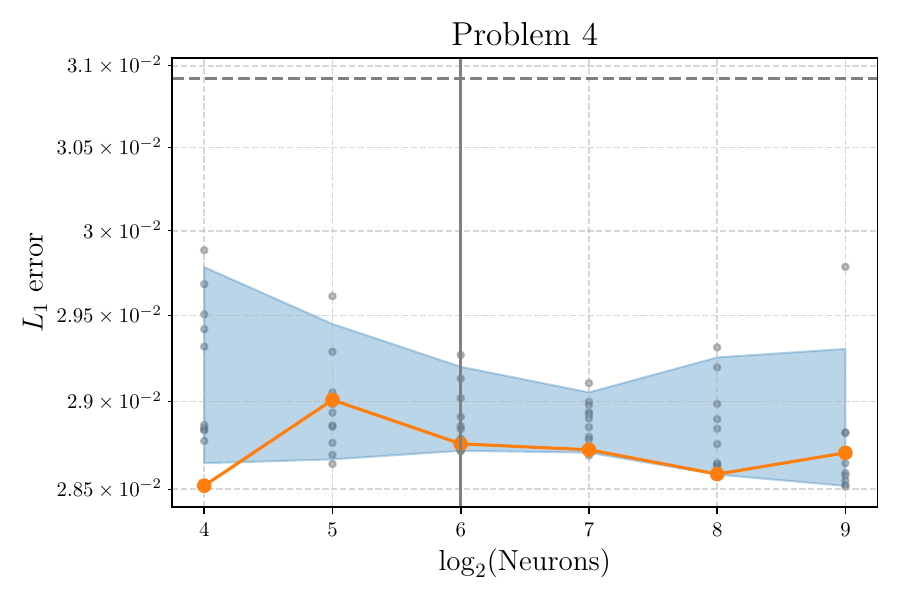}
    \end{subfigure}

    \caption{$L_1$ density error across problems for networks with two hidden layers of varying widths, rarefaction solvers always comprising half as many neurons as corresponding $g_p$ MLPs. All networks are trained on $10^{17}$ samples. Seed-to-seed variability is shown as a shaded region, while the main seed (42) is shown with full opacity. 
    Seed variability is relatively small across all model sizes, and it is minimal for models with $32$ to $128$ neurons per layer. The only exception is Problem 2, where, as in the previous subsection, a small number of seeds consistently fail to reproduce the solution. We again attribute this behavior to the large pressure jump in the initial conditions. 
    The vertical solid line marks the layer width adopted in the main text ($64$ neurons).}
    \label{fig:l1_modelsize}
\end{figure}

Figure~\ref{fig:l1_modelsize} shows the $L_1$ density error across the four shock tube problems for networks with two hidden layers of varying widths. Rarefaction networks always have half as many neurons per layer than the corresponding $g_p$ networks. All networks are trained on $10^{17}$ samples.
The main seed (42) is shown as a solid line, with the $5^{\rm th}$ to $95^{\rm th}$ percentile range across all seeds shown as a shaded region. The vertical solid line marks the layer width size in the main text ($64$ neurons).

Overall, the neural solver is found to be robust across a broad range of model sizes. As for the dataset size, the $L_1$ error variability is largest for very small and very large models, and remains overall under control for networks with $32$ to $128$ neurons per layer. 

In particular, Problem 4 shows consistently smaller errors than both HLLE and HLLC across all model sizes and seeds. Problem 1, on the other hand, has a larger spread of errors, but all seeds outperform the approximate solvers for layers of sizes between $32$ and $128$ neurons. For Problem 3 the errors are consistently quite small, and while the trends are not clearly monotonic, the largest observed spread is on the order of a factor $2$ for the largest models ($N_{\rm neurons} \geq 128$). Problem 2 remains the sole exception to this, with no clear trend in the errors across model sizes. As pointed out in the previous subsection, this behavior is likely due to the extreme pressure jump in the initial conditions, which places it near the boundary of the synthetic data distribution. Interestingly, in this case the errors are consistently smaller for larger models, suggesting that the neural solver could benefit from a larger $g_{\rm SR}$ network. Despite this, it is important to note that the relatively small amount of random seeds considered here ($11$) may not be sufficient to draw strong conclusions about the model size dependence of the neural solver.

\section{Improved robustness through model ensembling} \label{appendix:ensemble}
All the neural solvers considered so far comprised five individually trained compact neural networks, each specialized in solving a root-finding problem as generally formulated in Eq.~\eqref{eq:rootfinding_general}. While the networks generally manage to learn the features of the target function to a high degree of accuracy, the results of Appendix~\ref{appendix:dataset_size} show that a degree of randomness inherent to the training process persists in the simulation outcomes obtained with the neural solver. Moreover, increasing the model size (see Appendix~\ref{appendix:model_size}) appears to exacerbate this variability rather than mitigate it, likely due to the increased difficulty of training larger neural networks. Nevertheless, the average performance across seeds remains consistently good, and crucially, the majority of trained models achieve better accuracy than traditional approximate Riemann solvers.

This observation suggests a possible path towards reducing the variance in simulation results induced by randomness in model training: similar to standard practice in machine learning~\cite{Dietterich2000}, it is possible to construct an ensemble composed of several small networks and combine their predictions to produce the final fluxes. An obvious approach would be to average the predictions, but this risks poor performance if one or more networks in the ensemble are poorly conditioned on the inputs of interest. Alternative approaches could include voting schemes, where each network returns a confidence score, or weighted averages where the weights themselves are learned. However, in the present case, as discussed in the main text, a simple and inexpensive proxy for the prediction quality is readily available: the target function residual.

Following this intuition, in this section we explore the viability of an enhanced approach where each network in the solver is replaced by an ensemble of small MLPs, and the final prediction at each point is selected as the output with the lowest residual.

\begin{figure}[htbp]
    \centering

    % First row
    \begin{subfigure}[b]{0.45\textwidth}
        \centering
        \includegraphics[width=\textwidth]{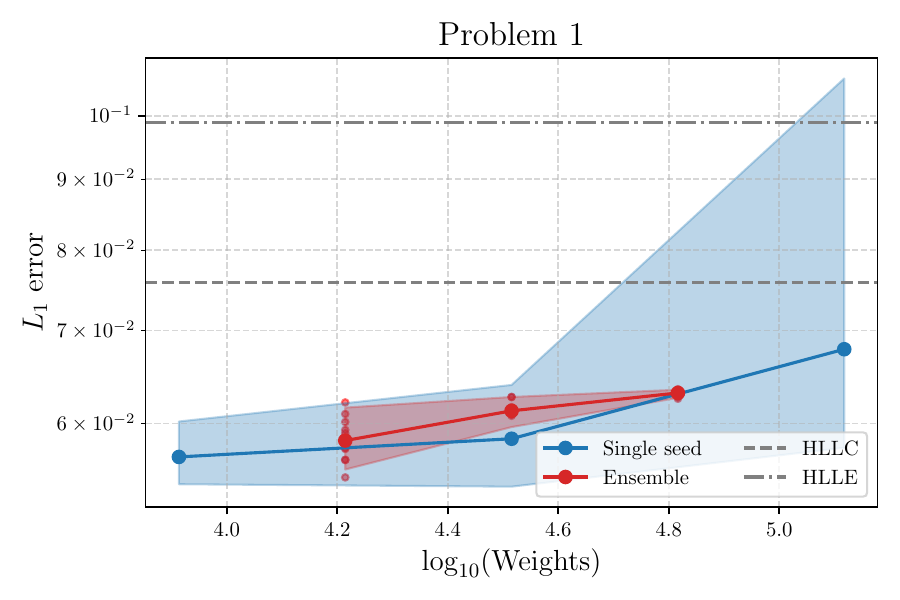}

    \end{subfigure}
    \hfill
    \begin{subfigure}[b]{0.45\textwidth}
        \centering
        \includegraphics[width=\textwidth]{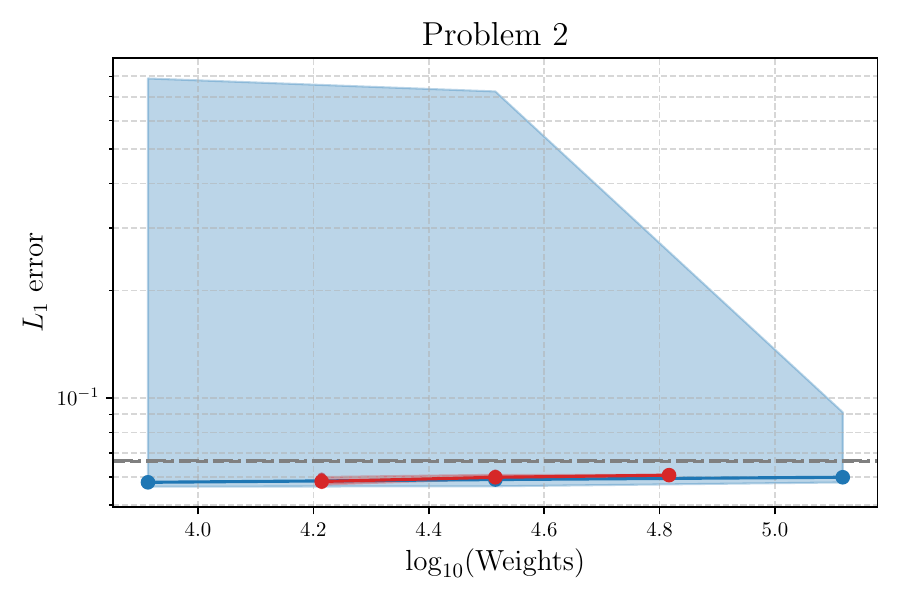}
    \end{subfigure}

    \vspace{1em}

    % Second row: centered with horizontal spacing

    \begin{subfigure}[b]{0.45\textwidth}
        \centering
        \includegraphics[width=\textwidth]{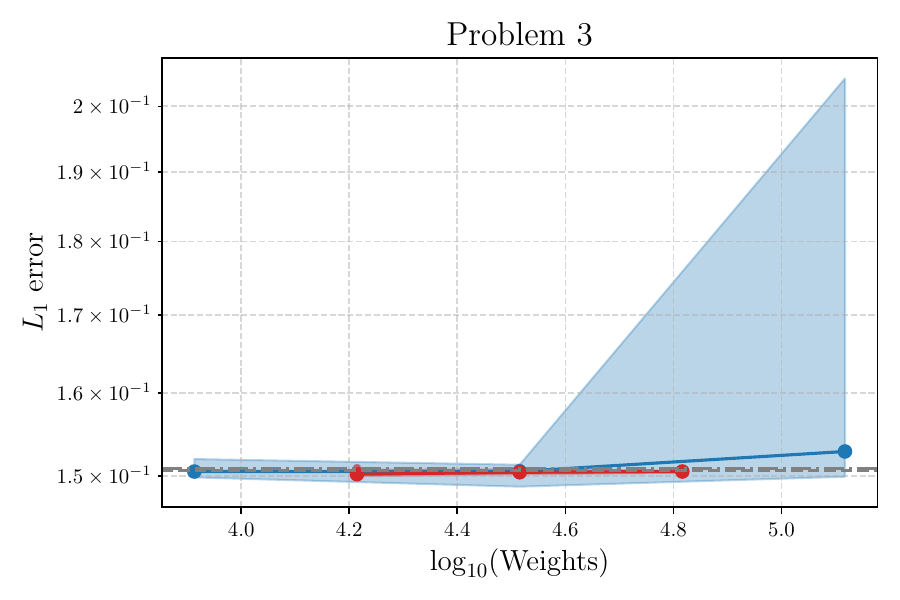}
    \end{subfigure}
    \hfill % gap between the two figs
    \begin{subfigure}[b]{0.45\textwidth}
        \centering
        \includegraphics[width=\textwidth]{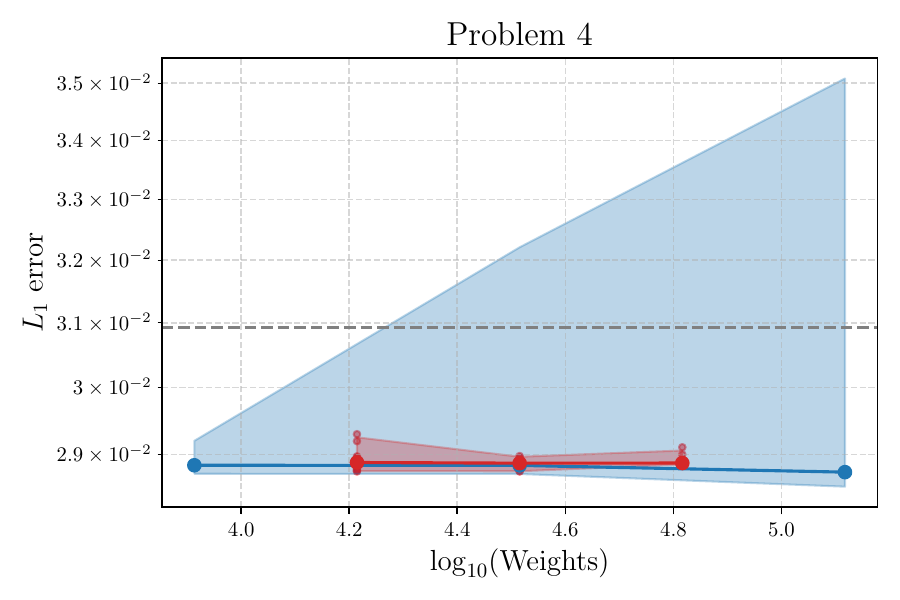}
    \end{subfigure}

    \caption{Comparison of $L_1$ density error across problems for single seed networks and equivalent ensembles. The ensembles are composed of $2,4$ and $8$ networks of $64$ neurons per layer, and are compared to appropriately scaled single-seed networks of $128,256$ and $512$ neurons per hidden layers. Median results for $11$ single-seed networks and $10$ ensembles are shown as solid lines, with $5^{\rm th}$ and $95^{\rm th}$ percentiles shown as shaded regions. Individual results from the ensembles are overlaid as red dots, and errors from approximate Riemann solvers are shown as horizontal dashed and dash-dotted lines.}
    \label{fig:l1_ensemble}
\end{figure}

Fig.~\ref{fig:l1_ensemble} shows the $L_1$ density errors across the four shock tube problems from the main text. Simulations are performed using a first-order scheme on a grid of 800 zones. We consider ensembles consisting of two, four, and eight networks, each with two hidden layers of 64 neurons, trained on $2^{17}$ samples. These results are compared to single-seed networks consisting of two hidden layers with 64, 128, and 256 neurons, trained on datasets of $2^{17}$, $2^{18}$, and $2^{19}$ samples, respectively. This ensures that both the number of learnable parameters and the total number of training samples are comparable across the two approaches. For each ensemble size, ten seed combinations are randomly selected from the set considered in Appendix~\ref{appendix:dataset_size}, whereas the single-seed networks are constructed with the same eleven seeds used in the previous sections.

In Fig.~\ref{fig:l1_ensemble}, median $L_1$ errors for both approaches are shown as solid lines (blue for single-seed solvers, red for ensembles) as a function of the number of weights in the networks' hidden layers (neglecting biases), with shaded regions indicating the 5$^{\mathrm{th}}$ to 95$^{\mathrm{th}}$ percentile ranges. A complete report of error statistics, including $1^{\rm st}$ and 99$^{\mathrm{th}}$ percentiles, is provided in Table~\ref{tab:ensemble_err}.

Both the figure and the table clearly indicate that, for a fixed effective model capacity (i.e., the number of learnable weights and the size of the training dataset), the ensemble approach consistently outperforms the corresponding single-seed solvers. In particular, ensemble models show a much narrower error distribution than the single network counterpart across all problems. Moreover, variance monotonically decreases for increasing model capacity in the ensemble solver approach, contrary to the single-seed case. 

The effectiveness of model ensembling is clearly showcased by the results for Problem 2, where outliers appeared in all other models considered so far, but are absent from all ensemble solvers. As can be observed from the Table, the $99^{\rm th}$ percentile of the $L_1$ error for the 2-net ensemble is one order of magnitude smaller than the corresponding percentile for the 128 neuron single-seed solver, and a factor of $2$ smaller than the $99^{\rm th}$ percentile of the 512 neuron solver. 

\begin{figure}
    \includegraphics[width=0.9\textwidth]{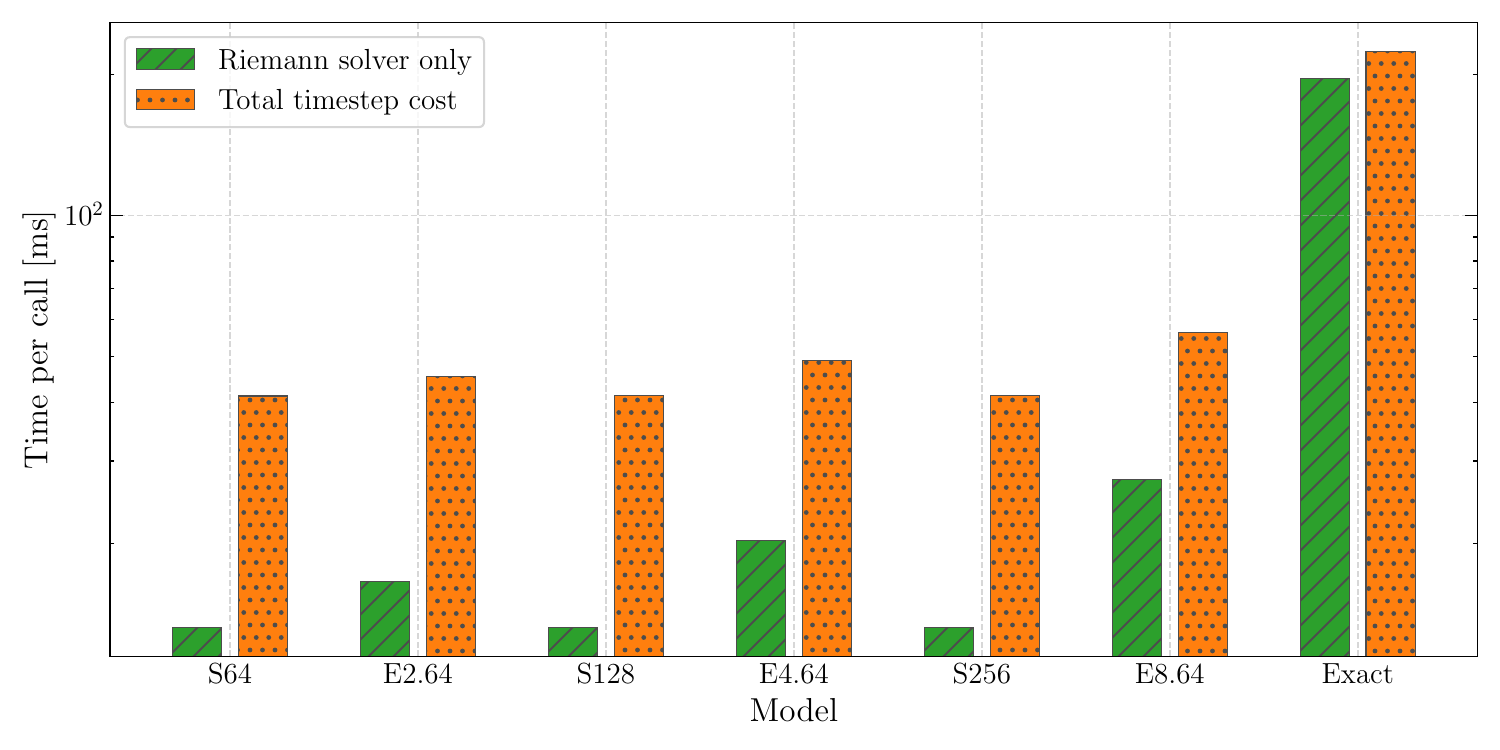}
    \caption{Riemann solver and total RHS computation runtime per call of single-seed and ensemble models across model capacity. }
    \label{fig:ensemble_timings}
\end{figure}

Naturally, model ensembling comes at a computational cost. Figure~\ref{fig:ensemble_timings} shows the computing time per call of the Riemann solver (green bars on the left) as well as for the full RHS evaluation (orange bars on the right) for single-seed models of varying capacity as well as equivalent ensemble models, together with the exact solver. It is important here to caution that the current implementation of ensembling is far from optimal, since inference is performed with each of the networks sequentially. This is particularly detrimental for the largest ensemble model, E8.64, which is a factor $2$ slower than the equivalent single-seed model. Despite this, the 2-net ensemble model, E2.64, is only about $25\%$ slower than even the single-seed 64 neuron solver, and even the slowest ensemble is a factor $7$ faster than the exact Riemann solver. Once again the effect on the total timestep time is far smaller. 

In conclusion, the ensemble prediction strategy appears to offer a viable path toward consistency and robustness for neural Riemann solvers in real-world applications. It is, however, important to keep in mind that in more complicated scenarios, where the function residual is not cheap to evaluate, a more sophisticated strategy to select the best prediction may be necessary.

\begin{table}[htbp]
    \centering
    \small
    \caption{Summary of \(L_1\) error statistics for different model configurations, evaluated at resolution 800. Single models are denoted as \(S\#\) (e.g., \(S128\) for a single network with 128 hidden neurons), while ensemble models are denoted as \(E{n}.\#\) (e.g., \(E2.64\) for an ensemble of two networks with 64 neurons each).}
    \label{tab:ensemble_err}
    \begin{tabular}{lccccc}
    \toprule
    \textbf{Problem 1} & & & & & \\
    \midrule
    \textbf{Model} & \textbf{Median} & \textbf{1\textsuperscript{st} \%} & \textbf{5\textsuperscript{th} \%} & \textbf{95\textsuperscript{th} \%} & \textbf{99\textsuperscript{th} \%} \\
    \midrule
    S64 & \(5.675\times 10^{-2}\) & \(5.410\times 10^{-2}\) & \(5.440\times 10^{-2}\) & \(5.862\times 10^{-2}\) & \(6.144\times 10^{-2}\) \\
    S128 & \(5.849\times 10^{-2}\) & \(5.345\times 10^{-2}\) & \(5.473\times 10^{-2}\) & \(6.308\times 10^{-2}\) & \(6.465\times 10^{-2}\) \\
    S256 & \(6.786\times 10^{-2}\) & \(5.650\times 10^{-2}\) & \(5.904\times 10^{-2}\) & \(9.110\times 10^{-2}\) & \(1.186\times 10^{-1}\) \\
    E2.64 & \(5.831\times 10^{-2}\) & \(5.500\times 10^{-2}\) & \(5.558\times 10^{-2}\) & \(6.159\times 10^{-2}\) & \(6.202\times 10^{-2}\) \\
    E4.64 & \(6.127\times 10^{-2}\) & \(5.890\times 10^{-2}\) & \(5.963\times 10^{-2}\) & \(6.269\times 10^{-2}\) & \(6.269\times 10^{-2}\) \\
    E8.64 & \(6.311\times 10^{-2}\) & \(6.253\times 10^{-2}\) & \(6.261\times 10^{-2}\) & \(6.345\times 10^{-2}\) & \(6.349\times 10^{-2}\) \\
    \midrule
    \textbf{Problem 2} & & & & & \\
    \midrule
    \textbf{Model} & \textbf{Median} & \textbf{1\textsuperscript{st} \%} & \textbf{5\textsuperscript{th} \%} & \textbf{95\textsuperscript{th} \%} & \textbf{99\textsuperscript{th} \%} \\
    \midrule
    S64 & \(5.794\times 10^{-2}\) & \(5.618\times 10^{-2}\) & \(5.651\times 10^{-2}\) & \(6.002\times 10^{-2}\) & \(1.371\times 10^{0}\) \\
    S128 & \(5.898\times 10^{-2}\) & \(5.641\times 10^{-2}\) & \(5.653\times 10^{-2}\) & \(1.852\times 10^{-1}\) & \(1.157\times 10^{0}\) \\
    S256 & \(5.986\times 10^{-2}\) & \(5.731\times 10^{-2}\) & \(5.862\times 10^{-2}\) & \(6.917\times 10^{-2}\) & \(1.087\times 10^{-1}\) \\
    E2.64 & \(5.825\times 10^{-2}\) & \(5.702\times 10^{-2}\) & \(5.710\times 10^{-2}\) & \(5.994\times 10^{-2}\) & \(6.003\times 10^{-2}\) \\
    E4.64 & \(5.985\times 10^{-2}\) & \(5.893\times 10^{-2}\) & \(5.924\times 10^{-2}\) & \(6.081\times 10^{-2}\) & \(6.119\times 10^{-2}\) \\
    E8.64 & \(6.068\times 10^{-2}\) & \(6.008\times 10^{-2}\) & \(6.016\times 10^{-2}\) & \(6.073\times 10^{-2}\) & \(6.073\times 10^{-2}\) \\
    \midrule
    \textbf{Problem 3} & & & & & \\
    \midrule
    \textbf{Model} & \textbf{Median} & \textbf{1\textsuperscript{st} \%} & \textbf{5\textsuperscript{th} \%} & \textbf{95\textsuperscript{th} \%} & \textbf{99\textsuperscript{th} \%} \\
    \midrule
    S64 & \(1.505\times 10^{-1}\) & \(1.498\times 10^{-1}\) & \(1.499\times 10^{-1}\) & \(1.517\times 10^{-1}\) & \(1.522\times 10^{-1}\) \\
    S128 & \(1.505\times 10^{-1}\) & \(1.478\times 10^{-1}\) & \(1.500\times 10^{-1}\) & \(1.511\times 10^{-1}\) & \(1.515\times 10^{-1}\) \\
    S256 & \(1.529\times 10^{-1}\) & \(1.490\times 10^{-1}\) & \(1.510\times 10^{-1}\) & \(1.552\times 10^{-1}\) & \(2.437\times 10^{-1}\) \\
    E2.64 & \(1.502\times 10^{-1}\) & \(1.500\times 10^{-1}\) & \(1.500\times 10^{-1}\) & \(1.507\times 10^{-1}\) & \(1.509\times 10^{-1}\) \\
    E4.64 & \(1.504\times 10^{-1}\) & \(1.501\times 10^{-1}\) & \(1.502\times 10^{-1}\) & \(1.508\times 10^{-1}\) & \(1.509\times 10^{-1}\) \\
    E8.64 & \(1.505\times 10^{-1}\) & \(1.503\times 10^{-1}\) & \(1.504\times 10^{-1}\) & \(1.506\times 10^{-1}\) & \(1.506\times 10^{-1}\) \\
    \midrule
    \textbf{Problem 4} & & & & & \\
    \midrule
    \textbf{Model} & \textbf{Median} & \textbf{1\textsuperscript{st} \%} & \textbf{5\textsuperscript{th} \%} & \textbf{95\textsuperscript{th} \%} & \textbf{99\textsuperscript{th} \%} \\
    \midrule
    S64 & \(2.884\times 10^{-2}\) & \(2.872\times 10^{-2}\) & \(2.872\times 10^{-2}\) & \(2.913\times 10^{-2}\) & \(2.926\times 10^{-2}\) \\
    S128 & \(2.884\times 10^{-2}\) & \(2.868\times 10^{-2}\) & \(2.876\times 10^{-2}\) & \(2.904\times 10^{-2}\) & \(3.474\times 10^{-2}\) \\
    S256 & \(2.874\times 10^{-2}\) & \(2.849\times 10^{-2}\) & \(2.858\times 10^{-2}\) & \(3.091\times 10^{-2}\) & \(3.843\times 10^{-2}\) \\
    E2.64 & \(2.888\times 10^{-2}\) & \(2.875\times 10^{-2}\) & \(2.876\times 10^{-2}\) & \(2.925\times 10^{-2}\) & \(2.929\times 10^{-2}\) \\
    E4.64 & \(2.887\times 10^{-2}\) & \(2.875\times 10^{-2}\) & \(2.875\times 10^{-2}\) & \(2.897\times 10^{-2}\) & \(2.897\times 10^{-2}\) \\
    E8.64 & \(2.887\times 10^{-2}\) & \(2.885\times 10^{-2}\) & \(2.885\times 10^{-2}\) & \(2.906\times 10^{-2}\) & \(2.910\times 10^{-2}\) \\
    \midrule
    \bottomrule
    \end{tabular}
    \end{table}

\FloatBarrier
\end{appendix}
% TODO:
% Provide your bibliography here. You have two options:

% FIRST OPTION - write your entries here directly, following the example below, including Author(s), Title, Journal Ref. with year in parentheses at the end, followed by the DOI number.
%\begin{thebibliography}{99}
%\bibitem{1931_Bethe_ZP_71} H. A. Bethe, {\it Zur Theorie der Metalle. i. Eigenwerte und Eigenfunktionen der linearen Atomkette}, Zeit. f{\"u}r Phys. {\bf 71}, 205 (1931), \doi{10.1007\%2FBF01341708}.
%\bibitem{arXiv:1108.2700} P. Ginsparg, {\it It was twenty years ago today... }, \url{http://arxiv.org/abs/1108.2700}.
%\end{thebibliography}

% \bibliographystyle{SciPost_bibstyle} % Include this style file here only if you are not using our template

\nolinenumbers

\end{document}